\documentclass[preprint,showpacs,preprintnumbers,amsmath,amssymb,superscriptaddress,nofootinbib]{revtex4}
\usepackage{graphicx,color}
\usepackage{amsmath,amssymb}
\usepackage{url}
\usepackage{epstopdf}
\newcommand{\hs}{\hspace*{0.5cm}}

\newcommand{\be}{\begin{equation}}
\newcommand{\ee}{\end{equation}}
\newcommand{\bea}{\begin{eqnarray}}
\newcommand{\eea}{\end{eqnarray}}
\newcommand{\nn}{\nonumber}
\newcommand{\crn}{\nonumber \\}

\newcommand{\al}{\alpha}
\newcommand{\la}{\lambda}

\newcommand{\fr}{\frac}

\newcommand{\bc}{\begin{center}}
\newcommand{\ec}{\end{center}}

\newcommand {\ba}{\begin{array}}
\newcommand {\ea}{\end{array}}
\newcommand{\ben}{\begin{enumerate}}
\newcommand{\een}{\end{enumerate}}

\usepackage{bm}
\usepackage{dcolumn}

\begin{document}

\title{Decay of standard-model-like Higgs boson $h\rightarrow \mu\tau$ in a 3-3-1 model with inverse seesaw neutrino masses}

\author{T. Phong Nguyen}\email{thanhphong@ctu.edu.vn}

\affiliation{Department of Physics, Can Tho University,
	3/2 Street, Can Tho, Vietnam}

\author{T.Thuy Le}\email{lethuthuy09a@gmail.com}

\affiliation{People's security high school II, Ap Bac street, My Tho, Tien Giang, Vietnam}

\author{T.T. Hong}\email{tthong@agu.edu.vn}
\affiliation{Department of Physics, An Giang University, Ung Van Khiem Street, Long Xuyen, An Giang, Vietnam}

\affiliation{Department of Physics, Hanoi Pedagogical University 2, Phuc Yen, Vinh Phuc, Vietnam}
\author{L.T. Hue\footnote{Corresponding author}}\email{lthue@iop.vast.ac.vn}

\affiliation{Institute for Research and Development, Duy Tan University, Da Nang City, Vietnam}

\affiliation{Institute of Physics, Vietnam Academy of Science and Technology, 10 Dao Tan, Ba	Dinh, Hanoi, Vietnam}

\begin{abstract}
 By adding new gauge singlets of neutral leptons, the improved versions of the 3-3-1 models with right-handed neutrinos  have been recently introduced in order to explain recent experimental neutrino oscillation data through the inverse seesaw mechanism. We prove that these models predict promising signals of lepton-flavor-violating decays of the standard-model-like Higgs boson  $h^0_1\rightarrow \mu\tau,e\tau$, which are suppressed in the original versions.  One-loop contributions to these decay amplitudes  are introduced in the unitary gauge. Based on a numerical investigation, we find that the branching ratios of the  decays $h^0_1\rightarrow\mu\tau,e\tau$ can reach values of $10^{-5}$ in the regions of parameter space satisfying the current experimental data of the decay $\mu\rightarrow e\gamma$.  The value of $10^{-4}$ appears when the Yukawa couplings of leptons are close to the perturbative limit. Some interesting properties of these regions of parameter space are also discussed.
\end{abstract} 
\maketitle
 \section{\label{intro} Introduction}
 \allowdisplaybreaks
Signals of lepton-flavor-violating decays of the standard-model-like Higgs boson (LFVHDs) were investigated at the LHC \cite{exLFVHD} not very long after its discovery in 2012 \cite{Hfound}. So far, the most stringent limits on the branching ratios (Br) of these decays  are Br$(h\rightarrow \mu\tau,e\tau)<\mathcal{O}(10^{-3})$, from the CMS Collaboration using data collected at a center-of-mass energy of 13 TeV. The sensitivities  of the planned colliders for LFVHD searches are predicted to reach the order of $10^{-5}$ \cite{exsensitive}.

On the theoretical side, model-independent studies showed that the LFVHDs predicted from models beyond the standard model (BSM) are  constrained indirectly from  experimental data such as lepton-flavor-violating  decays of charged leptons (cLFV) \cite{LFVtherotical}. Namely, they are affected most strongly by the recent experimental bound on Br$(\mu\rightarrow e\gamma)$. Fortunately, large branching ratios of the decays $h\rightarrow \mu\tau,e\tau$ are still allowed  up to the order of $10^{-4}$. Also, LFVHDs have been widely investigated in many specific BSM models, where the decay rates were indicated to be  close to  the upcoming sensitivities of colliders, including nonsupersymmetric \cite{NonLFVSUSY,HueNPB16} and supersymmetric versions \cite{LFVSUSY}. Among them, the models based on the gauge symmetry $SU(3)_C\times SU(3)_L\times U(1)_X$ (3-3-1) contain rich lepton-flavor-violating (LFV) sources which may result in interesting cLFV phenomenology such as charged lepton decays $e_i\rightarrow e_j\gamma$ \cite{ISS331,muega331,mega331b,mega331c}. In particular, it was shown that Br$(\mu\rightarrow e\gamma)$ is large in these models, and hence  it must be taken into account to constrain the parameter space. In addition, such rich LFV resources may give large LFVHD rates as promising signals of new physics.


Although the 3-3-1 models were introduced a long time ago \cite{old331,b331RHN}, LFVHDs have been investigated only in the  version with heavy neutral leptons assigned as  the third components of lepton (anti) triplets, where  active neutrino masses come from effective operators \cite{331LHN}. The  largest values of LFVHD rates were shown to be  $\mathcal{O}(10^{-5})$, originating from heavy neutrinos and charged Higgs bosons \cite{HueNPB16}. Improved versions consisting of new neutral lepton singlets were recently introduced \cite{ISS331,ISS331a}. They are more interesting because  they successfully explain the experimental neutrino data through the inverse seesaw (ISS) mechanism. We call them the 331ISS models for short. They  predict a large cLFV decay rate of $\mu\rightarrow e\gamma$ corresponding to recent experimental bounds. They may also contain dark matter candidates  \cite{ISS331,ISS331a}. These properties make them much more attractive than the original versions of 3-3-1 models with right-handed neutrinos (331RHN) \cite{b331RHN}. They predict suppressed LFV decay rates, because all neutrinos including exotic ones are extremely light.  Furthermore, loop corrections to the neutrino mass matrix must be taken into account to obtain an active neutrino mass spectrum that explains the experimental data \cite{chang}. Hence, LFV signals are an interesting way to distinguish the 331ISS and 331RHN models.  More specifically, a simple ISS extension of the SM  allows large Br$(h\rightarrow\mu\tau,e\tau)\sim \mathcal{O}(10^{-5})$ in the allowed regions satisfying Br$(\mu\rightarrow e\gamma)<4.2\times 10^{-13}$ \cite{issE}. Inspired by this, we will address the following questions in this work: how large is the Br$(h\rightarrow \mu\tau,e\tau)$ predicted by the 331ISS models under the experimental constraints  of the cLFV decays? and, are these branching ratios larger than the values calculated in the simplest ISS extension of the SM? Because these 331 models contain many more particles that contribute to LFV processes through loop corrections, either constructive or destructive correlations among them will strongly affect the allowed regions of the parameter space satisfying the current bound of the decay rate $\mu\rightarrow e\gamma$.  The most interesting allowed regions  will also allow large LFVHD rates,  which we will try to look for in this work. Because the discussion on the decay $h\rightarrow e\tau$ is rather similar to  the decay $h\rightarrow \mu\tau$, we only briefly mention the later.

Our paper is organized as follows. In Sec.~\ref{model} we  discuss the necessary ingredients of a 331ISS model for studying LFVHDs and how the ISS mechanism works to generate active neutrino parameters consistent with current experimental data. In Sec.~\ref{coupling} we present all couplings needed to determine the one-loop amplitudes of the LFVHDs of the SM-like Higgs boson. Sec.~\ref{numerical} we show important numerical LFVHD results predicted by the 331ISS model. Section .~\ref{conclusion} contains our conclusions. Finally, the Appendix  
lists all of the analytic formulas expressing one-loop contributions calculated in the unitary gauge.

\section{\label{model} The 331ISS model for tree-level neutrino masses}
\subsection{The  model and neutrino masses from the inverse seesaw mechanism}
First, we will consider a 331ISS model based on the original  331RHN model given in Ref. \cite{chang}, where  active neutrino masses and oscillations are generated from the ISS  mechanism. The quark sector and $SU(3)_C$ representations are irrelevant in this work, and hence they are omitted here.  The electric charge operator corresponding to  the gauge group  $SU(3)_L\times U(1)_X$ is $Q=T_3-\frac{1}{\sqrt{3}}T_8+X$, where $T_{3,8}$ are diagonal $SU(3)_L$ generators. Each  lepton family consists of  a  $SU(3)_L$ triplet $\psi_{aL}= (\nu_a,~e_a, N_a)_L^T\sim (3,-\frac{1}{3})$ and a right-handed charged lepton  $e_{aR}\sim (1,-1)$ with $a=1,2,3$.  Each left-handed neutrino $N_{aL}=(N_{aR})^c$  implies a new right-handed neutrino beyond the SM.  The three Higgs triplets are     $\rho=(\rho^+_1,~\rho^0,~\rho^+_2)^T\sim (3,\frac{2}{3})$,
$\eta=(\eta_1^0,~\eta^-,\eta^0_2)^T\sim (3,-\frac{1}{3})$, and $\chi=(\chi_1^0,~\chi^-,\chi^0_2)^T\sim (3,-\frac{1}{3})$.  The necessary vacuum expectation values for generating all tree-level quark masses are $\langle\rho \rangle=(0,\,\frac{v_1}{\sqrt{2}},\,0)^T$, $\langle \eta \rangle=(\frac{v_2}{\sqrt{2}},\,0,\,0)^T$ and $\langle \chi \rangle=(0,\,0,\,\frac{w}{\sqrt{2}})^T$.
 Gauge bosons in this model get masses through the covariant kinetic term of the Higgs bosons,
 \bea \mathcal{L}^{H}=\sum_{H=\chi,\eta,\rho} \left(D_{\mu}H\right)^{\dagger}\left(D_{\mu}H\right),\nn\eea
where the covariant derivative for the electroweak symmetry is defined as
\begin{eqnarray}
D_\mu  &=& \partial _\mu  - i g  {W}_\mu ^a{T^a} - {g_X}{T^9}X{X_\mu }, \, a=1,2,..,8, \label{cderivative}
\end{eqnarray}
and $T^9 \equiv \frac{I_3}{\sqrt{6}}$ and $\frac{1}{\sqrt{6}}$ for (anti)triplets and singlets \cite{buras}. It can be identified that
\be  g=e\, s_W, \hs \frac{g_X}{g}= \frac{3\sqrt{2}s_W}{\sqrt{3-4s^2_W}},  \ee
where $e$ and $s_W$ are, respectively, the electric charge and sine of the Weinberg angle, $s^2_W\simeq 0.231$.

The model includes two pairs of singly charged gauge bosons, denoted as  $W^{\pm}$  and $Y^{\pm}$, defined as
\bea W^{\pm}_{\mu}&=&\frac{W^1_{\mu}\mp i W^2_{\mu}}{\sqrt{2}},\hs m_W^2=\frac{g^2}{4}\left(v_1^2+v_2^2\right),\crn
Y^{\pm}_{\mu}&=&\frac{W^6_{\mu}\pm i W^7_{\mu}}{\sqrt{2}},\hs m_Y^2=\frac{g^2}{4}\left(w^2+v_1^2\right). \label{singlyG}\eea
The bosons $W^{\pm}$ are identified with the SM ones, leading to $v_1^2+v_2^2\equiv v^2=(246 \mathrm{GeV})^2$.  In the remainder of the text, we will consider in detail the simple case $v_1=v_2=v/\sqrt{2}=\sqrt{2}m_W/g$ given in Refs. \cite{NinhHue,HueNPB16}.

The two global symmetries-namely  normal and new lepton numbers denoted respectively, as $L$ and $\mathcal{L}$  were introduced. They are related to each other by \cite{chang,tully}:  $L=\frac{4}{\sqrt{3}}T_8+\mathcal{L}$. The detailed values of nonzero lepton numbers $L$ and $\mathcal{L}$ are listed in Table \ref{LLnumber}.
\begin{table}[h]
	\begin{tabular}{cc}
		\begin{tabular}{ccccccccccc}
			\hline
			Fields& $N_L$ & $\nu_L$ & $e_L$& $e_R$&$\rho^+_2$ & $\eta^0_2$ &$\chi^0_1$ &$\chi^-$ \\
			\hline
			$L$ & -1&1&1&1 &-2 &-2&2& 2\\
			\hline
		\end{tabular}&\hs
		\begin{tabular}{cccccc}
			\hline
			Fields& $\chi$ & $\eta$ & $\rho$& $\psi_{aL}$&$e_{aR}$  \\
			\hline
			$\mathcal{L}$ & $\frac{4}{3}$&$\frac{2}{3}$&$\frac{2}{3}$&$\frac{1}{3}$ &$1$ \\
			\hline
		\end{tabular}
	\end{tabular}
	\caption{Nonzero lepton number $L$ (left) and $\mathcal{L}$ (right) of leptons and Higgs bosons in the 331RHN}\label{LLnumber}
\end{table}

All  tree-level lepton mass terms come from the following Yukawa part:
\be  \mathcal{L}^{\mathrm{Y}}_l =-h^e_{ab}\overline{\psi_{aL}}\rho e_{bR}+
h^{\nu}_{ab} \epsilon^{ijk} \overline{(\psi_{aL})_i}(\psi_{bL})^c_j\rho^*_k
+\mathrm{H.c.},\label{Yl}\ee
where $\epsilon^{ijk}$ is the antisymmetric tensor  $\epsilon^{123}=1$,  $(\psi_{aL})^c\equiv((\nu_{aL})^c, (e_{aL})^c, (N_{aL})^c)^T$, and $h^{\nu}$ is an antisymmetric matrix, $h^{\nu}_{ab}=-h^{\nu}_{ba}$.
The first term of Eq. (\ref{Yl}) generates charged lepton masses $m_a$ satisfying  $h^{e}_{ab}\equiv \sqrt{2}\delta_{ab}m_a/v_1$
 in order to  avoid  LFV processes at the tree level. The second term in Eq. (\ref{Yl}) is expanded as follows:
 \begin{align}
 h^{\nu}_{ab} \epsilon^{ijk} \overline{(\psi_{aL})_i}(\psi_{bL})^c_j\rho^*_k
 &= 2 h^{\nu}_{ab} \left[-\overline{e_{aL}} (\nu_{bL})^c\rho^-_2-\overline{\nu_{aL}} (N_{bL})^c\rho^{0*}+ \overline{e_{aL}} (\nu_{bL})^c\rho^-_1 \right], \label{2rdterm}
 \end{align}
  where we have used the equality $\overline{N_{aL}} (\nu_{bL})^c=\overline{\nu_{bL}} (N_{aL})^c$,... The second term on the left-hand side of Eq. (\ref{2rdterm}) contributes a Dirac neutrino mass term $-\mathcal{L}_{\rm{mass}}^{\nu}=\overline{\nu_L}\,m_D\, N_R+\rm{H.c.}$, where $\nu_L\equiv(\nu_{1L},\nu_{2L},\nu_{3L})^T$, $N_R\equiv( (N_{1L})^c, (N_{2L})^c,(N_{3L})^c)^T$, and $(m_D)_{ab}\equiv \sqrt{2}\,v_1 h^{\nu}_{ab}$.  The model can predict a neutrino mass spectrum that is consistent  with current neutrino data \cite{pdg2016} when loop corrections are included, where  all new neutrinos are very light \cite{chang}. As a result, they will give suppressed LFV decay rates.

Now we consider a 331ISS model as an extension of the above  331RHN model, where three right-handed neutrinos which are gauge singlets, $X_{aR}\sim (1,0)$, $a=1,2,3$ are added. Now tree-level neutrino masses and mixing angles arise from the ISS mechanism.  Requiring that $\mathcal{L}$ is only softly broken, the additional Yukawa part is
\be
-\mathcal{L}_{X_R}= Y_{ab}\overline{\psi_{aL}}\,\chi X_{bR}+\frac{1}{2} (\mu_{X})_{ab}\overline{(X_{aR})^c}X_{bR}+ \mathrm{H.c.},
\label{NRY}\ee
where  $\mu_X$ is a $3\times3$ symmetric matrix and $L(X_{aR})=\mathcal{L}(X_{aR})=-1$. The last term in Eq. (\ref{NRY}) is the only one that violates both $L$ and $\mathcal{L}$, and hence it can be assumed to be small, which is exactly the case in the ISS models. The first term  generates mass for heavy neutrinos, resulting in a large Yukawa coupling $Y_{ab}$ with $SU(3)_L$ Higgs triplets. In addition, the ISS  mechanism allows for large entries in the Dirac mass matrix $m_D$ originated from Eq. (\ref{Yl}), which is the opposite of the well-known requirement in the 331RHN model.

In the basis $\nu'_{L}=(\nu_{L}, N_{L}, (X_R)^c)^T$ and $(\nu'_{L})^c=((\nu_L)^c, (N_L)^c, X_R)^T$,  Eqs. (\ref{Yl}) and   (\ref{NRY}) give a neutrino mass term corresponding to a  block form of the mass matrix, namely,
\bea -{L}^{\nu}_{\mathrm{mass}}=\frac{1}{2}\overline{\nu'_L}M^{\nu}(\nu'_L)^c +\mathrm{H.c.}, \,\mathrm{ where }\quad M^{\nu}=\begin{pmatrix}
	0	& m_D &0 \\
	m^T_D	&0  & M_R \\
0& M_R^T& \mu_X
\end{pmatrix},  \label{Lnu1}\eea
where   $M_R$  is  a $3\times3$ matrix  $(M_R)_{ab}\equiv Y_{ab}\frac{w}{\sqrt{2}}$ with $a,b=1,2,3$. Neutrino sub-bases are denoted as  $\nu_{R}=((\nu_{1L})^c,(\nu_{2L})^c,(\nu_{3L})^c)^T$, $N_R=((N_{1L})^c,(N_{2L})^c,(N_{3L})^c)^T$, and  $X_L=((X_{1R})^c,(X_{2R})^c,(X_{3R})^c)^T$.

The matrix $M^{\nu}$ can be written in the normal seesaw form,
\be  M^{\nu}=\begin{pmatrix}
	0& M_D \\
	M_D^T& M_N
\end{pmatrix}, \; \mathrm{where} \, M_D \equiv(m_D,\, 0),\;  \mathrm{and} \; M_N=\begin{pmatrix}
	0& M_R \\
	M_R^T& \mu_X
\end{pmatrix}. \label{Mnuss}\ee
The mass matrix  $M^{\nu}$ is   diagonalized by a $9\times9$ unitary matrix $U^{\nu}$ \cite{issE,issthhx},
\bea U^{\nu T}M^{\nu}U^{\nu}=\hat{M}^{\nu}=\mathrm{diag}(m_{n_1},m_{n_2},..., m_{n_{9}})=\mathrm{diag}(\hat{m}_{\nu}, \hat{M}_N), \label{diaMnu} \eea
where $m_{n_i}$ ($i=1,2,...,9$) are  mass eigenvalues of the nine mass eigenstates $n_{iL}$ (i.e., physical states of  neutrinos), $\hat{m}_{\nu}=\mathrm{diag}(m_{n_1},\;m_{n_2},\;m_{n_3})$, and  $\hat{M}_N=\mathrm{diag}(m_{n_4},\;m_{n_5},...,\;m_{n_{9}})$. They correspond to the masses of the three active neutrinos  $n_{aL}$ ($a=1,2,3$) and six extra neutrinos $n_{IL}$ ($I=4,5,..,9$).
The relations between the flavor  and mass eigenstates are
\bea\nu'_L=U^{\nu*} n_L, \hs \mathrm{and} \; (\nu'_L)^c=U^{\nu}  (n_L)^c, \label{Nutrans}
\eea
where $n_L\equiv(n_{1L},n_{2L},...,n_{9L})^T$ and $(n_L)^c\equiv((n_{1L})^c,(n_{2L})^c,...,(n_{9L})^c)^T$.

A four-component (Dirac) spinor $n_i$ is  defined as    $n_i\equiv(n_{iL},\; (n_{iL})^c)^T=n^c_i=(n_i)^c$, where the  chiral components are $n_{L,i}\equiv P_{L}n_i$ and $n_{R,i}\equiv P_{R}n_i= (n_{L,i})^c$ with chiral operators  $P_{L,R}=\frac{1\pm\gamma_5}{2}$. Similarly, the  definitions  for the original neutrino states  are  $\nu_a \equiv (\nu_{L,a},\; (\nu_{L,a})^c)^T$, $\nu_a \equiv (N_{L,a},\; (N_{L,a})^c)^T$,  $X_I\equiv((X_{R,I})^c,\; X_{R,I})^T$, and $\nu'=(\nu,\,N)^T$. The relations in Eq. (\ref{Nutrans}) can be written as follows:
\be P_L\nu'_i=\nu'_{i,L} =U^{\nu*}_{ij}n_{jL},\; \mathrm{and}\; P_R\nu'_i=\nu'_{iR} =U^{\nu}_{ij}n_{jR}, \hs i,j=1,2,...,9. \label{Nutrans2}\ee

In general,  $U^{\nu}$ is written in the form   \cite{numixing}
\be U^{\nu}= \Omega \left(
\begin{array}{cc}
	U & \mathbf{O} \\
	\mathbf{O} & V \\
\end{array}
\right), \hs
\label{Unuform}\ee
where  $\mathbf{O}$ is a  $3\times 6$ null matrix, and $U$, $V$, and $\Omega$  are   $3\times3$, $6\times 6$, and $9\times 9$ unitary matrices, respectively. $\Omega$  can be formally written as
\be \Omega=\exp\left(
\begin{array}{cc}
	\mathbf{O} & R \\
	-R^\dagger & \mathbf{O} \\
\end{array}
\right)=
\left(
\begin{array}{cc}
	1-\frac{1}{2}RR^{\dagger} & R \\
	-R^\dagger &  1-\frac{1}{2}R^{\dagger} R\\
\end{array}
\right)+ \mathcal{O}(R^3),
\label{Ommatrix}\ee
where $R$ is a $3\times 6$ matrix with the maximal absolute values for all entries $|R|$ satisfying $|R|<1$.  The matrix $U=U_{\mathrm{PMNS}}$ is the Pontecorvo-Maki-Nakagawa-Sakata (PMNS) matrix \cite{upmns},
\bea
U_{\mathrm{PMNS}}=\left(
\begin{array}{ccc}
	c_{12}c_{13} & s_{12}c_{13} & s_{13} e^{-i\delta} \\
	-s_{12}c_{23}-c_{12}s_{23}s_{13}e^{i\delta} & c_{12}c_{23}-s_{12}s_{23}s_{13}e^{i\delta} & s_{23}c_{13} \\
	s_{12}s_{23}-c_{12}c_{23}s_{13}e^{i\delta} & -c_{12}s_{23}-s_{12}c_{23}s_{13}e^{i\delta} & c_{23}c_{13} \\
\end{array}
\right) \mathrm{diag}(1,\; e^{i\frac{\alpha}{2}},\;e^{i\frac{\beta}{2}}),
\label{umns}\eea
and $c_{ab}\equiv\cos\theta_{ab}$, $s_{ab}\equiv\sin\theta_{ab}$. The  Dirac phase $\delta$ and Majorana phases $\alpha,\beta$ are fixed as $\delta=\pi,\alpha=\beta=0$. In the normal hierarchy scheme, the best-fit  values of  neutrino oscillation parameters are given as  \cite{pdg2016}
\bea \Delta m^2_{21}&=& 7.370\times 10^{-5}\;\mathrm{ eV^2},\hs  \Delta m^2= 2.50\times 10^{-3}\; \mathrm{eV^2},\crn
s^2_{12}&=&0.297,\; s^2_{23}=0.437,\; s^2_{13}=0.0214, \label{nuosc}\eea
where $ \Delta m^2_{21}=m^2_{n_2}-m^2_{n_1}$ and $\Delta m^2=m^2_{n_3}-\frac{\Delta m^2_{21}}{2}$.
The condition  $v_1 \ll w$ gives  the reasonable condition $|M_D|\ll |M_N|$, where $|M_D|$ and  $|M_N|$ denote the characteristic scales of $M_D$ and $M_N$. Hence, the following seesaw relations are valid  \cite{numixing}:
\bea   R^* &\simeq& \left(-m_DM^{-1}, \quad  m_D(M_R^T)^{-1}\right), \label{Rs}\\
  m_DM^{-1} m^T_D&\simeq&m_{\nu}\equiv U^*_{\mathrm{PMNS}}\hat{m}_{\nu}U^{\dagger}_{\mathrm{PMNS}}, \label{mnu}\\
V^* \hat{M}_N V^{\dagger}&\simeq& M_N+ \frac{1}{2}R^TR^* M_N+ \frac{1}{2} M_NR^{\dagger} R, \label{masafla}\eea
where
\be M\equiv M_R\mu_X^{-1}M_R^T. \label{deM}\ee
In the model under consideration, the Dirac neutrino mass matrix $m_D$ must be antisymmetric.  Equivalently, $m_D$ has only three independent parameters $x_{12},x_{13}$, and $z$,
\be m_D\equiv z \begin{pmatrix}
	0&x_{12}  &x_{13}  \\
	-x_{12}& 0 &1  \\
	-x_{13}& -1 &0
\end{pmatrix}, \label{mD1}\ee
where $z=\sqrt{2} v_1\,h^{\nu}_{23}$.
In contrast, the matrix $m_{\nu}$ in Eq. (\ref{mnu}) is symmetric, $(m_{\nu})_{ij}=(m_{\nu})_{ji}$, implying that 
$$0=(m_{\nu})_{ij}-(m_{\nu})_{ji}\sim x_{12}\left[(M^{-1})_{12}-(M^{-1})_{21}\right]+ x_{13}\left[(M^{-1})_{13}-(M^{-1})_{31}\right]+(M^{-1})_{23}-(M^{-1})_{32},$$
with $i,j=1,2,3$. This means that a symmetric matrix $M$ will give a right antisymmetric matrix $m_D$. To fit the neutrino data, there must exist  matrices $M$ and $m_D$ that satisfy the first equality in Eq. (\ref{mnu}).  Here we choose $M$ to be  symmetric for simplicity.  There must exist some sets of $z,x_{12},x_{13}$, and $M_{ij}$ ($i\le j\le3$) that satisfy the six equations  $\left(m_DM^{-1}m_D^T\right)_{ij}=(m_{\nu})_{ij}$, with $i\le j\le3$. From the three equations corresponding to $i=j=1,2,3$, we can write $(M^{-1})_{ii}$ as three functions of $z,x_{12}, x_{13},$ and $(M^{-1})_{ij}$ ($i\ne j$). Inserting them into the three remaining equalities, and taking some intermediate steps, we obtain
\bea -(m_{\nu})_{13} x_{12}+ (m_{\nu})_{12} x_{13}&=& (m_{\nu})_{11},\crn
-(m_{\nu})_{23} x_{12}+ (m_{\nu})_{22} x_{13}&=&(m_{\nu})_{12},\crn
-(m_{\nu})_{33} x_{12}+ (m_{\nu})_{23} x_{13}&=&(m_{\nu})_{13},\label{rmnuij}\eea
where we exclude the case of $x_{12},x_{13}=0$.
Solving the above three equations leads for  two  solutions of $x_{12,13}$ and a strict relation among $(m_{\nu})_{ij}$:
\bea x_{12}&=&\frac{(m_{\nu})_{11}(m_{\nu})_{23}-(m_{\nu})_{13}(m_{\nu})_{12}}{(m_{\nu})_{12}(m_{\nu})_{33}-(m_{\nu})_{13}(m_{\nu})_{23}},\crn
x_{13}&=&\frac{(m_{\nu})_{11}(m_{\nu})_{33}-(m_{\nu})^2_{13}}{(m_{\nu})_{12}(m_{\nu})_{33}-(m_{\nu})_{13}(m_{\nu})_{23}},\crn
0&=&(m_{\nu})_{11}(m_{\nu})^2_{23}+(m_{\nu})_{22}(m_{\nu})^2_{13}+ (m_{\nu})_{33}(m_{\nu})^2_{12}\crn
&-&(m_{\nu})_{11}(m_{\nu})_{22}(m_{\nu})_{33}- 2(m_{\nu})_{12}(m_{\nu})_{13}(m_{\nu})_{23}.
 \label{nxij}\eea
Interestingly, the last relation in Eq. (\ref{nxij}) allows us to predict possible values of the unknown neutrino mass based on the identification given in Eq. (\ref{mnu}). Using the experimental data given in Eq.  (\ref{nuosc}), we derive that $m_{\nu_1}=0$ in the normal hierarchy scheme.  The Dirac matrix now only depends on $z$:
\bea  m_D\simeq z\times \begin{pmatrix}
0	& 0.545 & 0.395 \\
-0.545	& 0 & 1 \\
	-0.395& -1 &0
\end{pmatrix}. \label{nmD}\eea
The above discussion also gives $M=\mathrm{diag}\left( 10^{10}z^2,\; 7.029\times10^{10} z^2 ,\;-2.377\times10^{11}z^2\right)$ for a  diagonal $M_R$. In this work, we also consider the simple case where $M_R$ is diagonal and all elements are positive. We also assume that $|m_{\nu}|< \mu_{X}\ll |m_D|< |M_R|$. We then derive that heavy neutrino masses are approximately equal to the entries of $M_R$, as given in Eq. (\ref{deM}). However, this approximation is not good for investigating LFVHDs, where a divergent cancellation in the numerical computation is strictly required. Instead, we will use the numerical solutions of heavy neutrino masses as well as the mixing matrix $U^{\nu}$ so that a total divergent part vanishes in  the final numerical results. This treatment will  avoid unphysical contributions originated from divergent parts.

Another parameterization  shown in Ref. \cite{ISS331}, can be applied to the  general cases of nonzero $\delta$ as well as  both the inverse and normal hierarchy cases of active neutrino masses. With the aim of finding regions with large LFVHDs, we will choose the simple case of $m_D$ given in Eq. (\ref{nmD}).

For simplicity in the numerical study, we will consider the diagonal matrix $M_{R}$ in  the degenerate case $M_R=M_{R_1}=M_{R_2}=M_{R_3}\equiv k\times z$.  The parameter $k$  will be fixed at small values that result in large LFVHD effects.  The total neutrino mass matrix in Eq.  (\ref{Lnu1}) depends on only the free parameter $z$.  The heavy neutrino masses and the matrix $U^{\nu}$ can be solved numerically, which is not affected by $z$ because $|\mu_{X}|\ll z$.

Using the exact numerical solutions for the neutrino masses and mixing matrix $U^{\nu}$ for our investigation, we emphasize that  the masses and mixing parameters of active neutrinos derived from the numerical diagonalization of the matrix $M^{\nu}$ given in Eq. (\ref{Lnu1}) should satisfy the $3\sigma$ constraint of the experimental data. In contrast, neutrino masses and mixing parameters defining the matrix $m_{\nu}$ in Eq. (\ref{mnu}), which are used to calculate the matrix $m_D$, are considered as free parameters. In other words, the experimental values of neutrino masses and mixing parameters are only used to estimate the allowed ranges of free parameters determining the mass matrix $M^{\nu}$.  After that, it is diagonalized numerically to find the neutrino masses as well as the mixing matrix $U^{\nu}$. The mixing parameters will be calculated from the matrix  $U_{\mathrm{PMNS}}$, which is related to $U^{\nu}$ by the relation (\ref{Unuform}). Requiring that the expansion of $\Omega$ in Eq. (\ref{Ommatrix})  and the  ISS condition $|\mu_X|> m_{n_3}$ are valid, we   find that small values of $k>1$ are allowed. In particular, we find that if three mixing parameters are fixed at the three respective center values, the two inputs for the  active neutrino masses may be outside of (but very close to) the $3\sigma$ ranges with $k=5$. When $k\ge5.5$, we always find that the input lies within the $3\sigma$ ranges of the experimental data that produces  the consistent numerical solutions of active neutrino masses. When $k\ge9$, the input corresponding to all center values given in Eq. (\ref{nuosc}) always produces  numerical solutions lying in the $3\sigma$ ranges of experimental data.

The LFVHD rates depend strongly on the  unitarity of the mixing matrix $U^{\nu}$ and heavy neutrino masses. On the other hand, they are  weakly affected by  the requirement that solutions for active neutrino masses and mixing parameters  satisfy the $3\sigma$ experimental data.  Hence, we will use the matrix $m_D$ given in Eq. (\ref{nmD}) and $k\ge 5.5$ for our numerical investigation. We  numerically checked that our choice produces  reasonable values for the  neutrino data  close to the $3\sigma$ ranges mentioned above.

 \subsection{Higgs and gauge bosons}
To study the LFVHD  effects, we will choose the simple case of the Higgs potential discussed in Refs. \cite{NinhHue,HueNPB16}, namely,
\bea
\mathcal{V} &=& \mu_1^2 \left( \rho^\dagger \rho + \eta^\dagger \eta \right) + \mu_2^2 \chi^\dagger \chi + \lambda_1 \left( \rho^\dagger \rho + \eta^\dagger \eta \right)^2 + \lambda_2 \left( \chi^\dagger \chi \right)^2 + \lambda_{12}  \left( \rho^\dagger \rho + \eta^\dagger \eta \right)  \left( \chi^\dagger \chi \right)
\crn
&&- \sqrt 2 f \left(\varepsilon_{ijk} \eta^i \rho^j \chi^k + \mathrm{H.c.}  \right),
\label{potential}
\eea
where $f$ is a mass parameter and is assumed to be real.
The detailed calculations for finding the masses and the mass eigenstates of Higgs bosons were presented in Refs.  \cite{NinhHue,HueNPB16}, where the minimum condition results in $v_1=v_2$. Here we will only list the part that is involved in LFVHDs.

The model contains two pairs of singly charged Higgs bosons $H^{\pm}_{1,2}$ and Goldstone bosons of the  gauge bosons $W^{\pm}$ and $Y^{\pm}$, which are denoted as $G^{\pm}_W$ and $G^{\pm}_Y$, respectively. The masses of all charged Higgs bosons are $m^2_{H^{\pm}_1}=f w (t_{\theta}^2 +1)$, $m^2_{H^{\pm}_2}=2 f w$, and $m^2_{G^{\pm}_{W}}=m^2_{G^{\pm}_{Y}}=0$, where $t_{\theta}=v_2/w$. The relations between the original and mass eigenstates of the charged Higgs bosons  are
\begin{eqnarray}
\left( \begin{array}{c}
\rho_1^\pm \\
\eta^\pm
\end{array} \right)= \dfrac{1}{\sqrt 2}\left( \begin{array}{cc}
-1 & 1\\
1 & 1
\end{array} \right) \left(\begin{array}{c}
G_W^\pm
\\ H_2^\pm
\end{array} \right), \hs \left( \begin{array}{c}
\rho_2^\pm \\
\chi^\pm
\end{array} \right) = \left( \begin{array}{cc}
- s_\theta & c_\theta \\
c_\theta & s_\theta
\end{array} \right) \left(\begin{array}{c}
G_Y^\pm
\\ H_1^\pm
\end{array} \right).
\label{EchargedH}
\end{eqnarray}
The neutral scalars are expanded as
\bea
 \rho^0 &=& \fr{1}{\sqrt{2}}(v_1 + S_1 + iA_1),\hs \eta^0_1 = \fr{1}{\sqrt{2}}(v_2 + S_2 + iA_2),  \quad
\chi^0_2 = \fr{1}{\sqrt{2}}(w + S'_3 + iA'_3),\crn
\eta^0_2 &=& \fr{1}{\sqrt{2}}(S'_2 + iA'_2), \quad
\chi^0_1 = \fr{1}{\sqrt{2}}(S_3 + iA_3). \label{Si}
\eea
There are four physical CP-even Higgs bosons $h^0_{1,2,3,4}$ and a Goldstone boson of the non-Hermitian gauge boson. The neutral Higgs components relevant for this work  are defined via
\bea
 \left(\begin{array}{c} S_1\\ S_2\\ S'_3 \end{array}\right) =
\left(
\begin{array}{ccc}
	-\fr{c_\alpha}{\sqrt{2}} & \fr{s_\alpha}{\sqrt{2}} & \fr{1}{\sqrt{2}} \\
	-\fr{c_\alpha}{\sqrt{2}} & \fr{s_\alpha}{\sqrt{2}} & -\fr{1}{\sqrt{2}} \\
	s_\alpha & c_\alpha & 0 \\
\end{array}
\right)\left(\begin{array}{c} h^0_1\\ h^0_2\\ h^0_3 \end{array}\right),
\label{mixing_CP_even_Higgs}
\eea
where $s_\alpha = \sin\alpha$ and $c_\alpha = \cos\alpha$, and they are defined by
\bea
s_\alpha &=& \fr{(4 \lambda_1-m^2_{h^0_1}/v_2^2)t_{\theta} }{r},\,
c_\alpha = \fr{\sqrt{2}\left(\lambda_{12}-\frac{f}{w}\right)}{r},\crn  r&=&\sqrt{2\left(\lambda_{12}-\frac{f}{w}\right)^2 + \left(4 \lambda_1 -m^2_{h^0_1}/v_2^2\right)^2 t_{\theta}^2}.
\label{angle_alpha}
\eea
There is one neutral CP-even Higgs boson $h^0_1$ with a mass proportional to the electroweak scale,
\be m^2_{h^0_1}=\frac{w^2}{2}\left[4\lambda_1 t_{\theta}^2 + 2\lambda_2 +\frac{ft_{\theta}^2}{w}-\sqrt{\left(2\lambda_2 +\frac{ft_{\theta}^2}{w}-4 \lambda_1 t^2_{\theta}\right)^2+8 t^2_{\theta}\left(\frac{f}{w}-\lambda_{12}\right)^2} \right]. \ee
The decoupling limit $t_{\theta}\ll 1$ ($v_1\ll w$) gives $m^2_{h^0_1}\sim \mathcal{O}(m_W^2)$ and  $s_{\alpha}\simeq0$ \cite{NinhHue}, resulting in the couplings similar to those predicted by the SM; see Table \ref{numbers}. Hence  $h^0_1$ is identified with the SM-like Higgs boson found at the LHC.

\section{\label{coupling}Couplings and analytic formulas  involved with LFVHDS}

\subsection{Couplings}
In this section we present Yukawa couplings in terms of $U^{\nu}$ and physical neutrino masses. From this, amplitudes and the LFVHD  rate are  formulated in terms of  physical masses and mixing parameters.  The equality derived from Eq. (\ref{diaMnu}),  $M^{\nu}=U^{\nu*}\hat{M}^{\nu}U^{\nu\dagger}$, gives
\bea
 M^{\nu}_{ab}&=&\left(U^{\nu*}\hat{M}^{\nu}U^{\nu\dagger}\right)_{ab}=0 \rightarrow U^{\nu*}_{ak}U^{\nu*}_{bk}m_{n_k}=0,\crn
\sqrt{2}v_1\,h^{\nu}_{ab} &=& (m_D)_{ab}= (M^{\nu})_{a(b+3)}=(U^{\nu*}\hat{M}^{\nu}U^{\nu\dagger})_{a(b+3)}=U^{\nu*}_{ak}U^{\nu*}_{(b+3)k}m_{n_k},\crn
 \frac{w}{\sqrt{2}} Y_{ab}&=&(M_R)_{ab}= (M^{\nu})_{(a+3)(b+6)}=U^{\nu*}_{(a+3)k}U^{\nu*}_{(b+6)k}m_{n_k},
\label{rel1}\eea
where $a,b=1,2,3$, and the sum is taken over $k=1,2,..,9$.

The relevant couplings in the first term of the Lagrangian (\ref{Yl}) are
\bea
&-& h^e_{ab}\overline{\psi_{aL}}\rho e_{bR}+{\rm h.c.}=- \frac{g  m_{a}}{ m_W}\left[ \overline{\nu_{aL}}e_{aR}\rho^+_1 +\overline{e_{aL}}e_{aR}\rho^0+ \overline{N_{aL}}e_{aR}\rho^+_2+\rm{ h.c.}\right]\crn
 &\supset&\frac{g\, m_{a}c_{\alpha}}{2m_W} h^0_1\overline{e_{a}}e_{a}-\frac{g\, m_{a}}{m_W}\left[c_{\theta}\left(U^{\nu}_{(a+3)i} \overline{n_{i}}P_Re_{a}H^+_1+ U^{\nu*}_{(a+3)i} \overline{e_{a}}P_Ln_{i}H^-_1\right)\right]\crn
 &&-\frac{g\, m_a}{\sqrt{2} m_W}\left[ \left(U^{\nu}_{ai} \overline{n_{i}}P_Re_{a}H^+_2+ U^{\nu*}_{ai} \overline{e_{a}}P_Ln_{i}H^-_2\right)\right].
\label{eephi}\eea
 The relevant couplings in the second term of the Lagrangian (\ref{Yl}) are
 \bea
 && h^{\nu}_{ab} \epsilon^{ijk} \overline{(\psi_{aL})_i}(\psi_{bL})^c_j\rho^*_k+ \rm{h.c.}\crn
 &=&2 h^{\nu}_{ab} \left[-\overline{e_{aL}} (\nu_{bL})^c\rho^-_2-\overline{\nu_{aL}} (N_{bL})^c\rho^{0*}+ \overline{e_{aL}} (\nu_{bL})^c\rho^-_1 \right]\crn
  &=&\frac{gc_{\alpha}}{2\,m_W}h^0_1\left[ \sum_{c=1}^3U^{\nu}_{ci}U^{\nu*}_{cj}\overline{ n_i}\left(m_{n_i}P_L+m_{n_j}P_R\right)n_j \right]\crn
 &-&\frac{gc_{\theta}}{m_W} \left[ (m_D)_{ab}U^{\nu}_{bi} H^-_1\overline{e_{a}}P_Rn_i+\rm{h.c.}\right]+\frac{g}{\sqrt{2}m_W} \left[ (m_D)_{ab}U^{\nu}_{(b+3)i} H^-_2\overline{e_{a}}P_Rn_i+\rm{h.c.}\right],
 \label{psipsiphi}\eea
 where the last line is derived following the calculation in Ref.~\cite{issthhx}: $\overline{ \nu_L}M_D((N_{L})^c,X_R)^T\leftrightarrow\overline{ \nu_{aL}}(M_D)_{aI} N_{IR}$.
 The first term in Eq. (\ref{NRY}) gives the following  couplings:
 \bea&-&Y_{ab}\overline{\psi_{aL}}\,\chi X_{bR}+\rm{h.c.}\crn
 &=&-\frac{\sqrt{2}}{w} (M_R)_{ab}\left[ \overline{\nu_{aL}}\chi^0_1 +\overline{e_{aL}}\chi^{-} + \overline{N_{aL}}\chi^0_2\right] X_{bR} +\rm{h.c.}\crn
 &\supset& -\frac{g t_{\theta}} {\sqrt{2}m_W}(M_R)_{ab}\left[s_{\alpha} U^{\nu}_{(a+3)i} U^{\nu}_{(b+6)j}\overline{n_{i}}P_Rn_{j}h^0_1+ \sqrt{2}s_{\theta}U^{\nu}_{(b+6)i} \overline{e_{a}}P_Rn_{i}H^-_1 +\rm{h.c.} \right], \label{NRYe}\eea
 where  we have used  $t_{\theta}=v_1/w\rightarrow1/w=t_{\theta}/v_1=gt_{\theta}/(\sqrt{2} m_W)$.
  The LFVHD couplings between leptons and charged gauge bosons   $(W^{\pm},~ Y^{\pm})$ are
 \bea  \mathcal{L}^{\ell\ell V}=\overline{\psi_{aL}}\gamma^{\mu}D_{\mu}\psi_{aL}
 &\supset&\frac{g}{\sqrt{2}} \left( \overline{e_{aL}}\gamma^\mu  \nu_{aL}W^{-}_{\mu} + \overline{e_{aL}}\gamma^\mu  N_{aL} Y^{-}_{\mu} \right)+\mathrm{H.c.}\crn
 &=&\frac{g}{\sqrt{2}} \left[ U^{\nu*}_{ai} \overline{e_{a}}\gamma^\mu P_L  n_{i}W^{-}_{\mu} + U^{\nu}_{ai} \overline{ n_{i}}\gamma^\mu P_L e_{a}W^{+}_{\mu}\right.\crn
 &+& \left.U^{\nu*}_{(a+3)i}\overline{e_{a}}\gamma^\mu  P_L n_{i} Y^{-}_{\mu} + U^{\nu}_{(a+3)i}\overline{n_{i} }\gamma^\mu  P_L e_{a}Y^{+}_{\mu} \right],\label{llv1}\eea
 where $D_{\mu}=\partial_{\mu}-\frac{ig}{2}\left(W^a_{\mu}\lambda^a+t \times(-\frac{1}{3})B_{\mu}\right)$, $\lambda_a$ ($a=1,2,..,8$) are the Gell-mann matrices, and $t=g_X/g$.  The charged gauge bosons are $W^{\pm}_{\mu}=\frac{W^1_{\mu}\mp i W^{2}_{\mu}}{\sqrt{2}}$ and $Y^{\pm}_{\mu}=\frac{W^6_{\mu}\pm i W^{7}_{\mu}}{\sqrt{2}}$.

 By defining a symmetric coefficient $\lambda^0_{ij}=\lambda^0_{ji}$, namely,
 \be \lambda^0_{ij}=\sum_{c=1}^3\left(U^{\nu}_{ci}U^{\nu*}_{cj}m_{n_i}+U^{\nu*}_{ci}U^{\nu}_{cj}m_{n_j}\right)-\sum_{c,d=1}^3
\sqrt{2} t_{\alpha}t_{\theta}(M^*_R)_{cd} \left[U^{\nu*}_{(c+3)i} U^{\nu*}_{(d+6)j}+U^{\nu*}_{(c+3)j} U^{\nu*}_{(d+6)i}\right],\crn
\ee
 the coupling $h^0_1\overline{n_i}n_j$ derived from Eqs. (\ref{psipsiphi}) and (\ref{NRYe}) is written in the symmetric form
 $\frac{gc_{\alpha}}{4m_W}h^0_1\overline{n_i}\left[\lambda^0_{ij}P_L+\lambda^{0*}_{ij}P_R\right]n_j$, which gives the  right vertex coupling based on the Feynman rules given in Ref. \cite{spinor}.
 The Yukawa couplings of charged Higgs bosons are defined by
 \bea \lambda^{R,1}_{ai}&=&m_{a}U^{\nu}_{(a+3)i},\quad
  \lambda^{L,1}_{ai}= \sum_{c=1}^3\left[(m_D^*)_{ac}U^{\nu*}_{ci}+ t^2_{\theta}(M_R^*)_{ac}U^{\nu*}_{(c+6)i}\right],\crn
   \lambda^{R,2}_{ai}&=&m_{a}U^{\nu}_{ai},\quad
   \lambda^{L,2}_{ai}= -\sum_{c=1}^3(m_D^*)_{ac}U^{\nu*}_{(c+3)i},.
   \eea
 Finally, all of the couplings involved in LFV processes are listed in Table~\ref{numbers}.
    \begin{table}[h]
   	\begin{tabular}{|c|c|}
   		\hline
   		Vertex & Coupling \\
   			\hline
   		$ h^0_1 \overline {e_a}e_a$ & $\frac{igm_{a}}{2m_W}c_{\alpha}$ \\
   			\hline
   		$ h^0_1 \overline {n_i}n_j$ & $\frac{igc_{\alpha}}{2m_W}\left(\lambda^0_{ij}P_L+\lambda^{0*}_{ij}P_R\right)$ \\
   		\hline
   		$  H_1^+\overline{n_i} e_b$, 	$H_1^-\overline {e_a} n_i $ & $-\frac{igc_{\theta}}{m_W}\left(\lambda^{L,1}_{bi}P_L+\lambda^{R,1}_{bi}P_R\right)$,  $-\frac{igc_{\theta}}{m_W}\left(\lambda^{L,1*}_{ai}P_R+\lambda^{R,1*}_{ai}P_L\right)$\\
   		\hline
   		$  H_2^+\overline{n_i} e_b$, 	$H_2^-\overline {e_a} n_i $ & $-\frac{ig}{\sqrt{2}m_W}\left(\lambda^{L,2}_{bi}P_L+\lambda^{R,2}_{bi}P_R\right)$,  $-\frac{ig}{\sqrt{2} m_W}\left(\lambda^{L,2*}_{ai}P_R+\lambda^{R,2*}_{ai}P_L\right)$\\	
   		\hline
   		$ W_\mu^+ \overline{ n_i} e_b$,  	$   W_\mu^-\overline {e_a}n_i$ & $\frac{ig}{\sqrt{2}}U^{\nu}_{bi}\gamma^\mu P_L$, $\frac{ig}{\sqrt{2}}U^{\nu*}_{ai}\gamma^\mu P_L$\\
   		\hline
   		$Y_\mu^+ \overline{n_i}  e_b$, 	$Y_\mu^-  \overline {e_a} n_i $& $\frac{ig}{\sqrt{2}}U^{\nu}_{(b+3)i}\gamma^\mu P_L$, $\frac{ig}{\sqrt{2}}U^{\nu*}_{(a+3)i}\gamma^\mu P_L$\\
   			\hline
 $H_1^+h^0_1Y_{\mu}^-$, $Y^+_{\mu}H_1^-h^0_1$ &$\dfrac{ig}{2 \sqrt 2} \left( c_\al c_\theta + \sqrt 2 s_\al s_\theta \right)  \left( p_{h_1^0} - p_{H_1^+} \right)^\mu$, $  \dfrac{ig}{2 \sqrt 2} \left( c_\al c_\theta + \sqrt 2 s_\al s_\theta \right)  \left( p_{H_1^-} - p_{h_1^0} \right)^\mu $  \\
   	 		\hline
   		$h^0_1 W^+_{\mu }W^-_{\nu}$ &  $ -ig m_W c_\alpha\,g^{\mu\nu} $\\
   	  		\hline
   	  		$h^0_1 Y^+_{\mu }Y^-_{\nu}$ & $ \frac{ig m_Y}{\sqrt 2} \left( \sqrt 2 s_\al c_\theta - c_\al s_\theta \right) g^{\mu\nu}$\\
   	  		\hline
   		$h^0_1 H^+_1H^-_1$ &$ i\lambda^{\pm}_{H_1}= - i w \left[ s_\al c_\theta^2 \lambda_{12} + 2 s_\al s_\theta^2 \lambda_2 - \sqrt 2 \left( 2 c_\al c_\theta^2 \lambda_1 + c_\al s_\theta^2 \lambda_{12}  \right) t_\theta - \dfrac{\sqrt 2}{v_3} f c_\al c_\theta s_\theta \right]$  \\
   		\hline
   		$ h^0_1H^+_2H^-_2$ &$i\lambda^{\pm}_{H_2}=-i v_1 \left(-2 \sqrt 2 c_\al \lambda_1 + \dfrac{s_\al v_3 \lambda_{12} + s_\al f}{v_1}  \right) $\\
   		\hline
   	\end{tabular}
   	\caption{Couplings related to the SM-like Higgs decay $h^0_1\rightarrow e_ae_b$ in the 331ISS model. All momenta in the Feynman rules corresponding  to these vertices are incoming.  \label{numbers}}
   \end{table}
The model predicts that the following couplings are zero:  $ h^0_1 W^{\pm}Y^{\mp}$, $ h^0_1 W^{\pm}H^{\mp}_{1,2}$, $h^0_1 Y^{\pm}H^{\mp}_{2},$ and $h^0_1 H^{\pm}_1H^{\mp}_2$.
\subsection{Analytic formulas}
\begin{figure}[h]
	\centering
	\includegraphics[width=14cm]{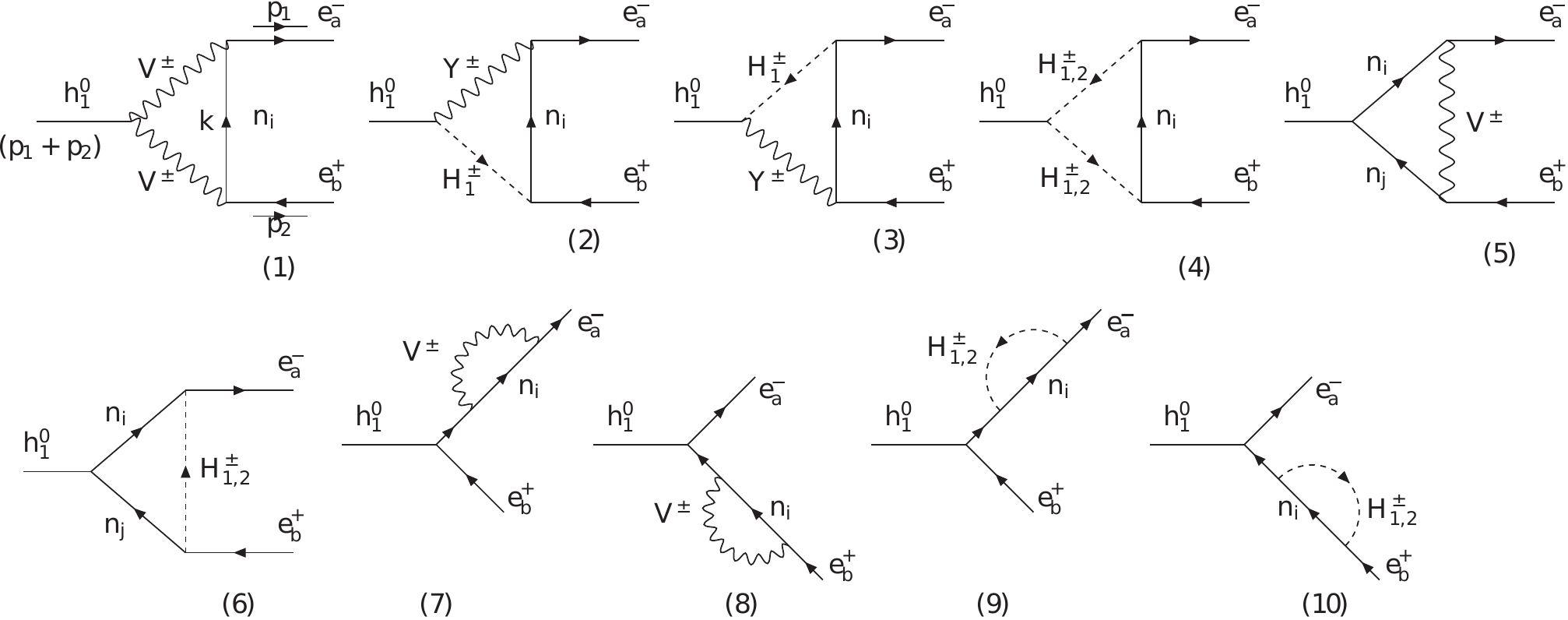}\\
	\caption{One-loop Feynman diagrams contributing to the decay $h^0_1\rightarrow e_a e_b$ in the unitary gauge.  Here $V^\pm=W^{\pm},\,Y^{\pm}$.}\label{hlilj1}
\end{figure}

The effective Lagrangian of the LFVHDs of the SM-like Higgs boson  $h^0_1\rightarrow e_a^{\pm}e_b^{\mp}$ is
$$ \mathcal{L}^{\mathrm{LFVH}}= h^0_1 \left(\Delta_{(ab)L} \overline{e_a}P_L e_b +\Delta_{(ab)R} \overline{e_a}P_R e_b\right) + \mathrm{H.c.},$$
where the scalar factors $\Delta_{(ab)L,R}$  arise from the loop contributions. In the unitary gauge, the one-loop Feynman diagrams contributing to this LFVHD amplitude are shown in Fig. \ref{hlilj1}.

The partial width of  the decay is
\be
\Gamma (h_1^0\rightarrow e_ae_b)\equiv\Gamma (h_1^0\rightarrow e_a^{-} e_b^{+})+\Gamma (h^0_1\rightarrow e_a^{+} e_b^{-})
=  \fr{ m_{h^0_1} }{8\pi }\left(\vert \Delta_{(ab)L}\vert^2+\vert \Delta_{(ab)R}\vert^2\right), \label{LFVwidth}
\ee
with the condition  $m_{h_1^0}\gg m_{a,b}$. Where $m_{a,b}$ are the masses of muon and tau, respectively. The on-shell conditions for external particles are $p^2_{1,2}=m_{a,b}^2$  and $ p_{h_1^0}^2 \equiv( p_1+p_2)^2=m^2_{h_1^0}$. The corresponding branching ratio is  Br$(h^0_1\rightarrow e_ae_b)= \Gamma (h_1^0\rightarrow e_ae_b)/\Gamma^{\mathrm{total}}_{h_1^0},$ where $\Gamma^{\mathrm{total}}_{h_1^0}\simeq 4.1\times 10^{-3}$ GeV \cite{pdg2016, Gah01}.  The $\Delta_{(ab)L,R}$ can be written as
\be \Delta_{(ab)L,R} =\sum_{i=1,5,7,8} \Delta^{(i)W}_{(ab)L,R} + \sum^{10}_{i=1} \Delta^{(i)Y}_{(ab)L,R},  \label{deLR}\ee
where the analytic forms of  $\Delta^{(i)W}_{(ab)L,R}$ and $\Delta^{(i)Y}_{(ab)L,R}$ are  shown in the Appendix. They can be calculated using the unitary gauge with the same techniques given in Refs. \cite{issthhx,HueNPB16}. We have crosschecked this with FORM \cite{form}.

The divergence cancellation in the total amplitude (\ref{deLR}) is proved analytically in the Appendix, based on the following strict equality:
 \begin{align}
 \label{fmnu2}
 U^{\nu}(\hat{M}^{\nu})^2U^{\nu\dagger}&=(U^{\nu*}\hat{M}^{\nu}U^{\nu \dagger})^*U^{\nu*}\hat{M}^{\nu}U^{\nu \dagger}=M^{\nu*}M^{\nu}\crn
 &=\begin{pmatrix}
 m_D^*m_D^T& 0 & m_D^*M_R \\
 0	& m_D^{\dagger}m_D+M_R^*M_R^T & M^*_R\mu_X \\
 M_R^{\dagger}m_D^T	& \mu^*_XM_R^T & M^\dagger_RM_R+\mu_X^*\mu_X
 \end{pmatrix}
 \end{align}
In the model under consideration, the divergent parts coming from the contributions of charged Higgs and heavy gauge bosons are related to both $(M^{\nu*}M^{\nu})_{(a+3)(b+3)}$  and $(M^{\nu*}M^{\nu})_{(a+6)(b+6)}$ ($a,b\leq 3$), which are affected by heavy neutrino masses.  The cancellation in the total divergent part  requires that the physical heavy neutrino masses and $U^{\nu}$ must be the exact values.  Hence, approximate forms of the heavy neutrino masses and neutrino mixing matrix derived from the ISS mechanism cannot be applied. In contrast,  we checked numerically that these  formulas  are safely used in the usual minimal ISS version  extended directly from the SM, because the divergent parts are only involved with the elements $(M^{\nu*}M^{\nu})_{ab}=(m_D^*m_D^T)_{(ab)}$.

 Many of the contributions listed in Eq. (\ref{deLR}) are suppressed, and hence they can be ignored in our numerical computation. From now on, we just focus on the decay $h^0_1\rightarrow\mu\tau$, and hence the simplified notations $\Delta_{L,R}\equiv\Delta_{(23)L,R}$ will be used. The decay $h^0_1\rightarrow\, e\tau$ has similar properties, so we do not need to discuss it more explicitly.  We can see that $|\frac{\Delta_L}{\Delta_R}|\simeq \mathcal{O}\left( \frac{m_{\mu}}{m_{\tau}}\right)$. In addition, we prove in the Appendix that the following combinations are finite: $\Delta_{L,R}^{(1+5)W}$, $\Delta_{L,R}^{(7+8)W}$, $\Delta_{L,R}^{(4)YH_2}$, $\Delta_{L,R}^{(6+9+10)YH_2}$, $\Delta_{L,R}^{(4)YH_1}$, $\Delta_{L,R}^{(7+8)Y}$, and   $(\Delta_{L,R}^{(1+2+3+5)Y} +\Delta_{L,R}^{(6+9+10)YH_1})$. With $m_{\mu,\tau}\ll m_W$, we have $B^{(1)}_1+B^{(2)}_1, B^{(2)}_1-B^{(2)}_0\simeq0$, and hence $\Delta_{L,R}^{(7+8)W},\Delta_{L,R}^{(7+8)Y}\simeq0$. The two contributions $\Delta_{L,R}^{(4)YH^{\pm}_{1,2}}$ are also suppressed with a large  $m_{H^{\pm}_2}$ for about a few TeV.

The four diagrams 4, 6, 9 and 10 in Fig. 1 include contributions from both charged Higgs bosons. They are not significantly affected by the $SU(3)_L$ scale $m_Y$, and thus they may enhance the partial decay widths of the LFVHDs if charged Higgs masses are small.

The regions of parameter space predicting large branching ratios for LFVHDs  are affected strongly by the current experimental bound Br$(\mu\rightarrow e\gamma)<4.2\times 10^{-13}$ \cite{muega2016}.  A very good approximate formula for this decay rate in the limit $m_{\mu},m_e\rightarrow0$ is  \cite{muega331}
\begin{equation}\label{brmuega}
\mathrm{Br}(\mu\rightarrow e\gamma)= \frac{12\pi^2}{G_F^2}|D_R|^2,
\end{equation}
where $G_F=g^2/(4\sqrt{2}m_W^2)$ and  $D_R$ is the one-loop contribution from charged gauge and  Higgs boson mediations, $D_R=D^{W}_R +D^{Y}_R +D^{H^{\pm}_1}_R +D^{H^{\pm}_2}_R$. The analytic forms are
\begin{align}
D^{W}_R&=-\frac{eg^2}{32\pi^2m_W^2} \sum_{i=1}^9U^{\nu*}_{ai}U^{\nu}_{bi}F(t_{iW}),\crn
D^{Y}_R&=  -\frac{eg^2}{32\pi^2m_Y^2} \sum_{i=1}^9U^{\nu*}_{(a+3)i}U^{\nu}_{(b+3)i}F(t_{iY}), \crn
D^{H^{\pm}_k}_R&=-\frac{eg^2f_k}{16\pi^2m_W^2} \sum_{i=1}^9\left[\frac{ \lambda^{L,k*}_{ai}  \lambda^{L,k}_{bi}}{m^2_{H^{\pm}_k}}\times\frac{1-6t_{ik} +3 t^2_{ik} +2t^3_{ik} -6t^2_{ik} \ln(t_{i_k})}{12 (t_{ik}-1)^4} \right.\crn
&+\left. \frac{m_{n_i} \lambda^{L,k*}_{ai} \lambda'^{R,k}_{bi}}{m^2_{H^{\pm}_k}}\times \frac{-1 +t_{ik}^2 -2t_{ik} \ln(t_{ik})}{2(t_{ik}-1)^3} \right],
\end{align}
where
\begin{align}
b&=2,\;a=1,\; t_{iW}\equiv \frac{m^2_{n_i}}{m_W^2},\; t_{iY}\equiv \frac{m^2_{n_i}}{m_Y^2}, \; t_{ik}\equiv \frac{m^2_{n_i}}{m^2_{H^{\pm}_k}},\crn
f_1&\equiv \frac{1}{2},\; f_2\equiv c^2_{\theta},\;  \lambda'^{R,1}_{bi}\equiv U^{\nu}_{(b+3)i},\;  \lambda'^{R,2}_{bi}\equiv U^{\nu}_{bi},\crn
F(x)&\equiv -\frac{10-43x+78x^2-49x^3 +4x^4 +18x^3\ln(x)}{12(x-1)^4}.
\end{align}
Because all charged Higgs bosons couple with heavy neutrinos through the Yukawa coupling matrix $h^{\nu}_{ab}$, this matrix is strongly affected by the upper bound $\mathcal{O}(10^{-13})$ on Br$(\mu\rightarrow e\gamma)$.  In fact, our numerical investigation shows that the allowed regions with light charged Higgs masses are very narrow. The previous investigation in Ref. \cite{ISS331} showed that the 331ISS models predicts a large Br$(\mu\rightarrow e\gamma)$, where the allowed regions discussed there were chosen such  that $k\sim \mathcal{O}(10^3)$ and $M_R\le 1$ TeV, implying that $z\sim \mathcal{O}(1)$ eV. We checked that our formulas are consistent with these results. In general,  the allowed regions are very strict, and satisfy  one of the following conditions. First, the regions have a small $z$ and large $|M_R|$ and $m_{H_2^{\pm}}$, implying $k\gg1$, including those mainly discussed in Ref. \cite{ISS331}. Second, the regions allow for a large $m_D$ and small $k$, but the strong destructive correlation between the two-loop contributions of charged  gauge and  Higgs bosons must happen.  These regions were also considered in Ref.~\cite{ISS331}, but they were not given much attention.  They are very interesting because they predict  large branching ratios for LFVHDs and  light particles such as new neutrinos and charged Higgs bosons, which could be found at the LHC and planned colliders \cite{Higg331,Heavynu}.  Hence, our numerical investigation will focus on this case.

\section{\label{numerical} Numerical discussion on LFVHDS}
\subsection{\label{setuppa} Setup parameters}
 To numerically investigate  the LFVHDs of the SM-like Higgs boson, we will use the following well-known experimental parameters  \cite{pdg2016}: the mass of the $W$ boson $m_{W}=80.385$ GeV, the charged lepton masses $m_e=5\times 10^{-4}$ GeV, $m_{\mu}=0.105$ GeV, and $m_{\tau}=1.776$ GeV, the SM-like Higgs mass $m_{h^0_1}=125.1$ GeV,  and the gauge coupling of the $SU(2)_L$ symmetry $g\simeq 0.651$.

Combined with the discussion in Sec.~\ref{model}, the independent parameters are the heavy neutrino mass  scale  $M_R=\mathrm{diag}(M_{R},\, M_{R},\, M_{R})$,  the heavy gauge boson mass $m_Y$ considered as the $SU(3)_L$ breaking scale, the charged Higgs boson mass $m_{H^{\pm}_2}$,  the characteristic scale of $m_D$  defined as the parameter  $z$, and the two Higgs self-couplings $\lambda_{1,12}$.

Other parameters can be calculated in terms of the above free ones, namely,
\begin{align}
 v_1 &=v_2= \frac{\sqrt{2}m_W}{g},\; s_{\theta}=\frac{m_W}{m_Y\sqrt{2}}, \;w=\frac{2 m_Y}{g\,c_{\theta}}, \; f=\frac{g\,c_{\theta}\,m^2_{H^{\pm}_2}}{4m_Y},\; m^2_{H^{\pm}_1}=\frac{m^2_{H^{\pm}_2}}{2}(t^2_{\theta} +1).   \label{nonfreeparameter}
\end{align}
 Apart from that, the mixing parameter $\alpha $ of  the neutral CP-even Higgs was defined in Eq. (\ref{angle_alpha}). The Higgs self-coupling $\lambda_{2}$ is determined as \cite{HueNPB16}
\begin{equation}\label{la2}
\lambda_{2}=\frac{t^2_{\theta}}{2}\left(\frac{m^2_{h^0_1}}{v_1} -\frac{m^2_{H^{\pm}_2}}{2 w^2}\right)+\frac{\left(\lambda_{12}-\frac{m^2_{H^{\pm}_2}}{2w^2}\right)^2}{4\lambda_1 -\frac{m^2_{h^0_1}}{v_1^2}}.
\end{equation}
In the model under consideration with the quark sector given in Refs. \cite{chang, Higg331}, only the charged Higgs bosons $H^{\pm}_2$ couple with all SM leptons and quarks.  They have been investigated  at the LHC in the direct production $pp\rightarrow t(b)H^{\pm}$, which then  decay into two final fermion states \cite{cHiggs1}. But the specific  constraints on them in the framework of the 3-3-1 models  have not  been mentioned yet, to the best of our knowledge. Instead, the lower bounds on their masses have been discussed recently  based on recent data of neutral meson mixing $B_0-\bar{B}_0$, where a reasonable lower bound of $m_{H^{\pm}_2}\geq480$ GeV was concerned \cite{Higg331}.

The values of $\lambda_{1,2,12}$ must satisfy  theoretical conditions of unitarity and the Higgs potential must be  bounded from below, as mentioned in Ref. \cite{HueNPB16}. The heavy charged gauge boson mass $m_Y$ is related to the recent lower constraint of neutral gauge boson $Z'$ in this model.

For the above reasons, the default values of the free parameters chosen for our numerical investigation are as follows. Without loss of generality, the Higgs self-couplings are fixed as $\la_1=1,\,\lambda_{12}=-1$, which also guarantee that all couplings of the SM-like Higgs boson approach  the SM limit when $t_{\theta}\rightarrow0$.  The default value $m_Y=4.5$ TeV satisfies all recent constraints \cite{Higg331,mYbound}. The parameter $z$ will be considered in the range of the perturbative limit $z < 2\sqrt{\pi}\times v_1\simeq 617$ GeV: in particular, we will fix $z=50,200,400$, $500$, and $600$ [GeV]. Finally, the charged Higgs mass  $m_{H^{\pm}_2}$ will be investigated mainly in the range of $300$ to $5\times 10^4$ GeV, where large values of LFVHDs may appear.

\subsection{\label{appro}Numerical results}
\begin{figure}[h]
	\centering
	\begin{tabular}{cc}
		\includegraphics[width=7cm]{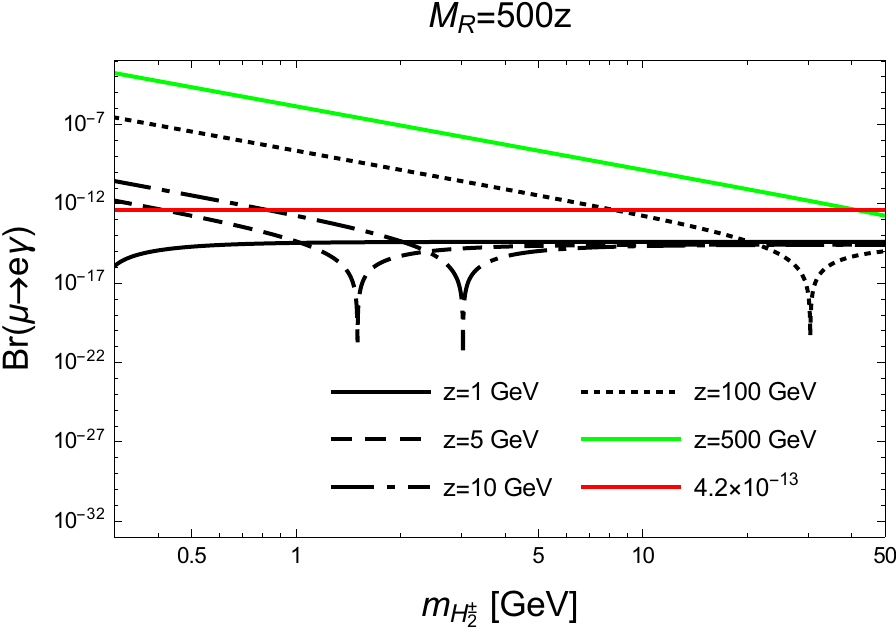}&
		\includegraphics[width=7cm]{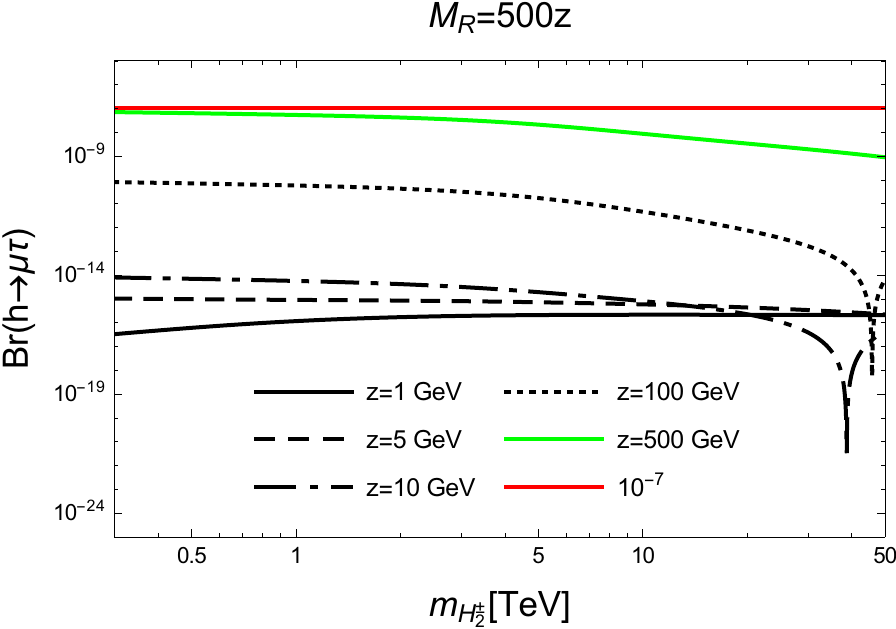}\\
		\end{tabular}
	\caption{Br$(\mu\rightarrow e\gamma)$ (left) and Br$(h^0_1\rightarrow \mu\tau)$ (right) as  functions of $m_{H^{\pm}_2}$ with $k=500$.}\label{MRggMd}
\end{figure}

First, we reproduce the regions  mentioned in Ref. \cite{ISS331}, where $M_R$ was chosen to be from hundreds of GeV to 1 TeV, and the scale of $m_D$ (namely, $z$), was a few GeV, corresponding to $k\gg1$.  As a result, the respective regions of parameter space always satisfy the experimental bound on Br$(\mu\rightarrow e\gamma)$ with large enough  $m_{H^{\pm}_2}$. These regions are shown in Fig. \ref{MRggMd} with fixed $z=1,5,10,100$, and $500$ GeV.
 All allowed regions (i.e, those that satisfy the upper bound Br$(\mu\rightarrow e\gamma)<4.2\times 10^{-13}$) give a small Br$(h^0_1\rightarrow\mu\tau)<\mathcal{O}(10^{-9})$. In general, for larger $k$ we checked numerically that the values of the branching ratio of LFVHDs will decrease significantly, and hence we will not discuss this further.

\begin{figure}[h]
\centering
\begin{tabular}{cc}
	\includegraphics[width=7cm]{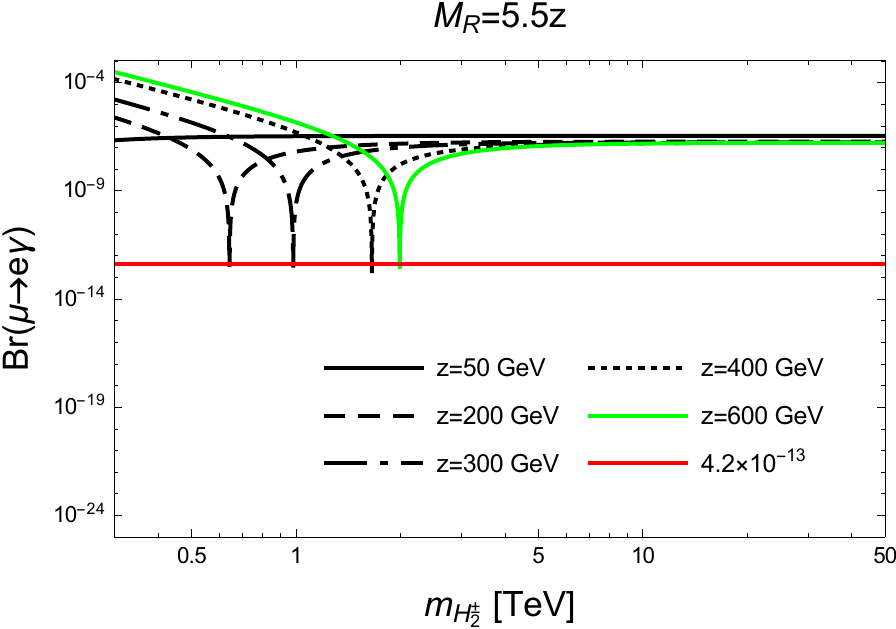}&
	  \includegraphics[width=7cm]{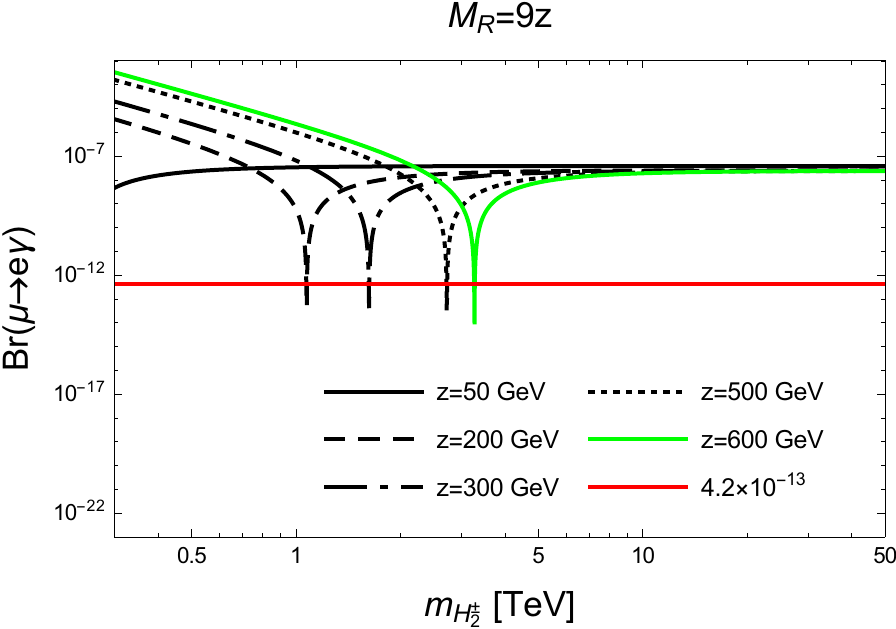}\\
	\includegraphics[width=7cm]{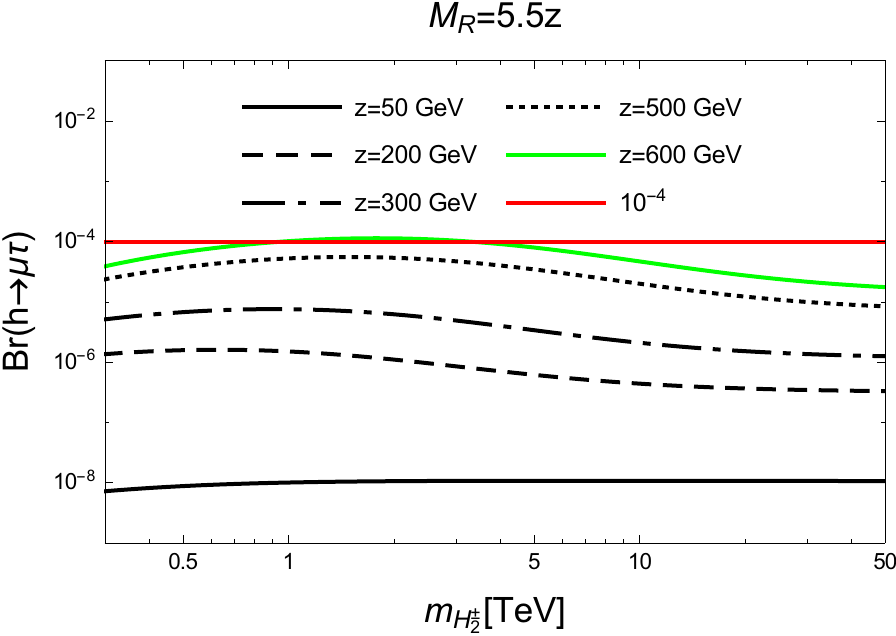}&
	\includegraphics[width=7cm]{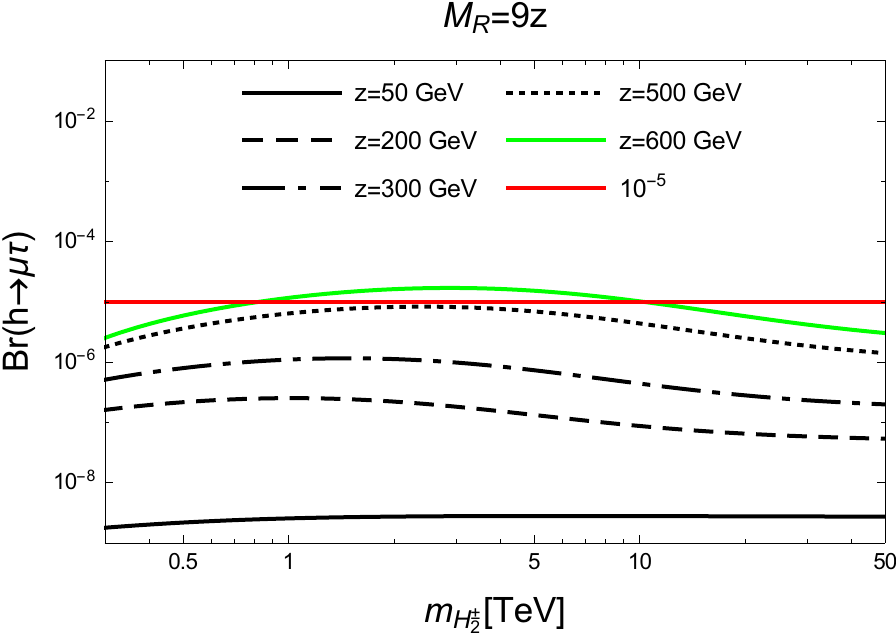}\\
\end{tabular}
\caption{Br$(\mu\rightarrow e\gamma)$ (upper) and Br$(h^0_1\rightarrow \mu\tau)$ (lower) as  functions of $m_{H^{\pm}_2}$ with $k=5.5$ (left) and $k=9$ (right).}\label{Fdemc2}
\end{figure}

With small values of $k=5.5$ and $9$, the dependence of both Br$(\mu\rightarrow e\gamma)$ and Br$(h\rightarrow\mu\tau)$ on $m_{H^{\pm}_2}$ with fixed $z$ are shown in Fig.~\ref{Fdemc2}.
Most  regions of the parameter space are ruled out by the bound on Br$(\mu\rightarrow e\gamma)$, except for narrow parts where particular contributions from charged Higgs and gauge bosons are destructive. This interesting property of the 331ISS model was indicated previously in Ref.~\cite{ISS331}. Furthermore, it predicts  allowed regions that give a large Br$(h^0_1\rightarrow \mu\tau)$. In particular,  the largest values can reach $\mathcal{O}(10^{-4})$ when $k=5.5$ and $z=600$ GeV, which is very close to the perturbative limit. In general, the illustrations in two Figs.~\ref{MRggMd} and~\ref{Fdemc2} suggest that this branching ratio is enhanced significantly for smaller $k$ and larger $z$, but changes slowly with the change of large $m_{H^{\pm}_2}$. In contrast, small $m_{H^{\pm}_2}$ plays a very important role in creating  allowed regions that predict a large LFVHD. Br$(\mu\rightarrow e\gamma)$ does not depend on $m_{H^{\pm}_2}$  when it  is large enough. Furthermore, the branching ratio decreases with increasing $k$ and it will go below the experimental bound if $k$ is large enough.

\begin{figure}[ht]
	\centering
	\begin{tabular}{cc}
		\includegraphics[width=7cm]{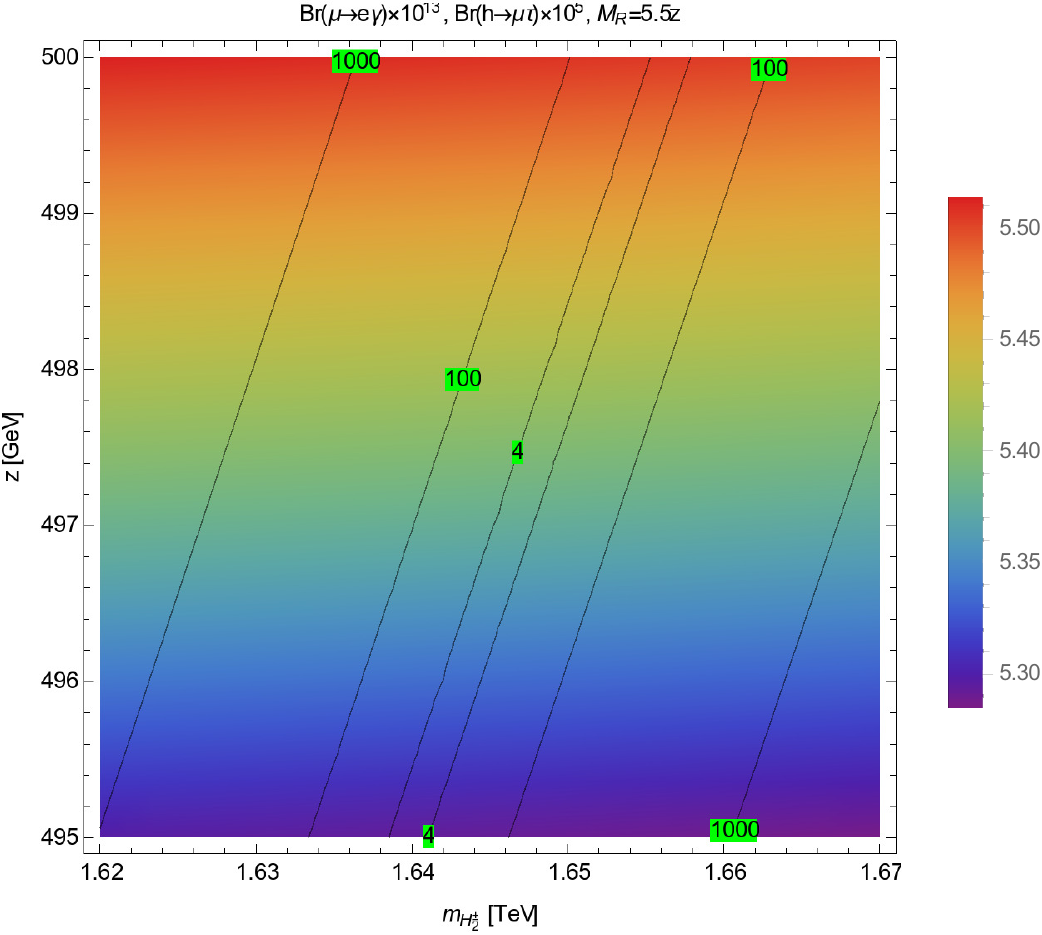}&
		\includegraphics[width=7cm]{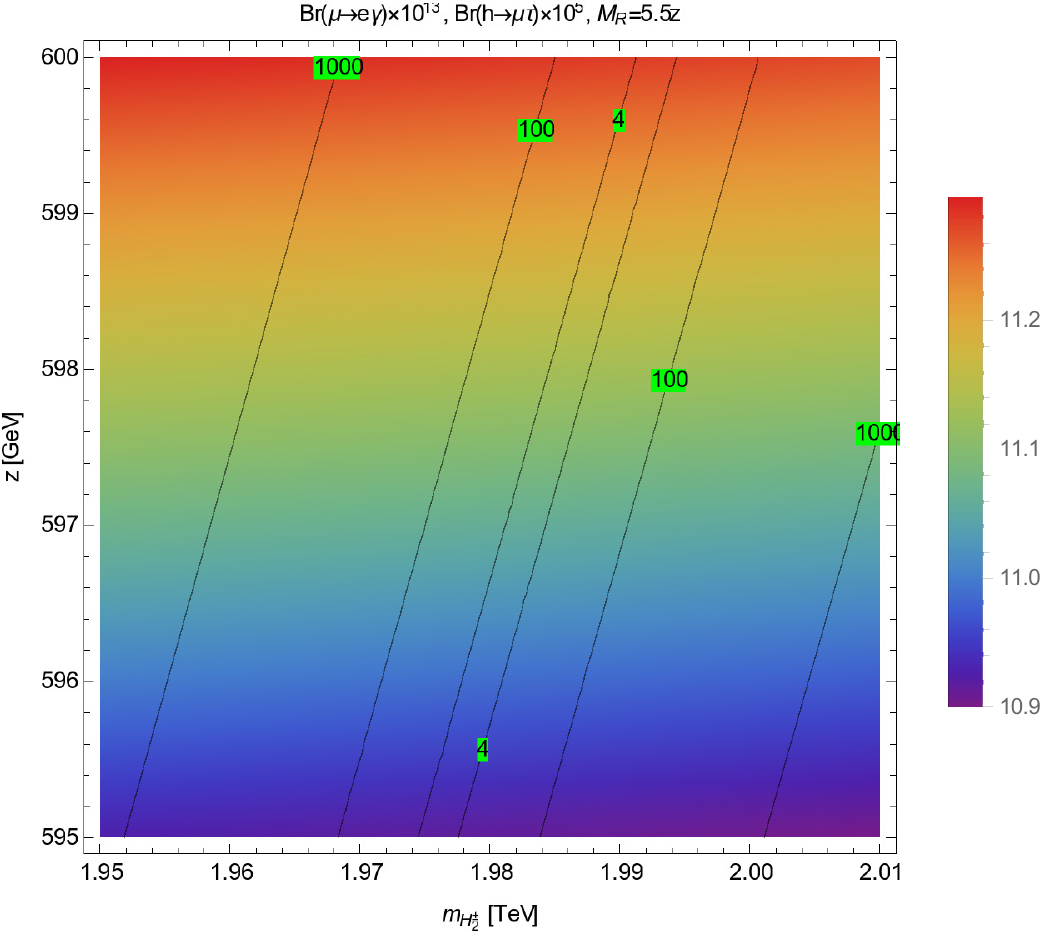}\\
		\includegraphics[width=7cm]{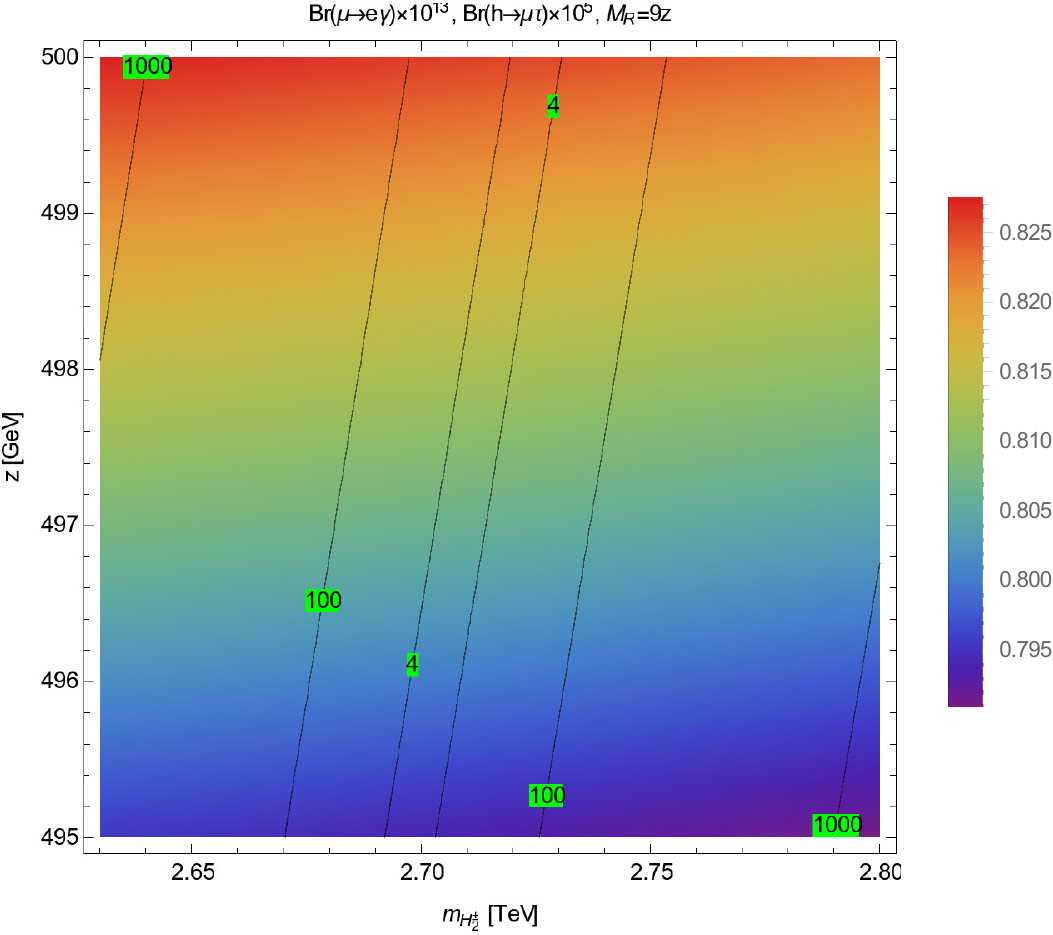}&
		\includegraphics[width=7cm]{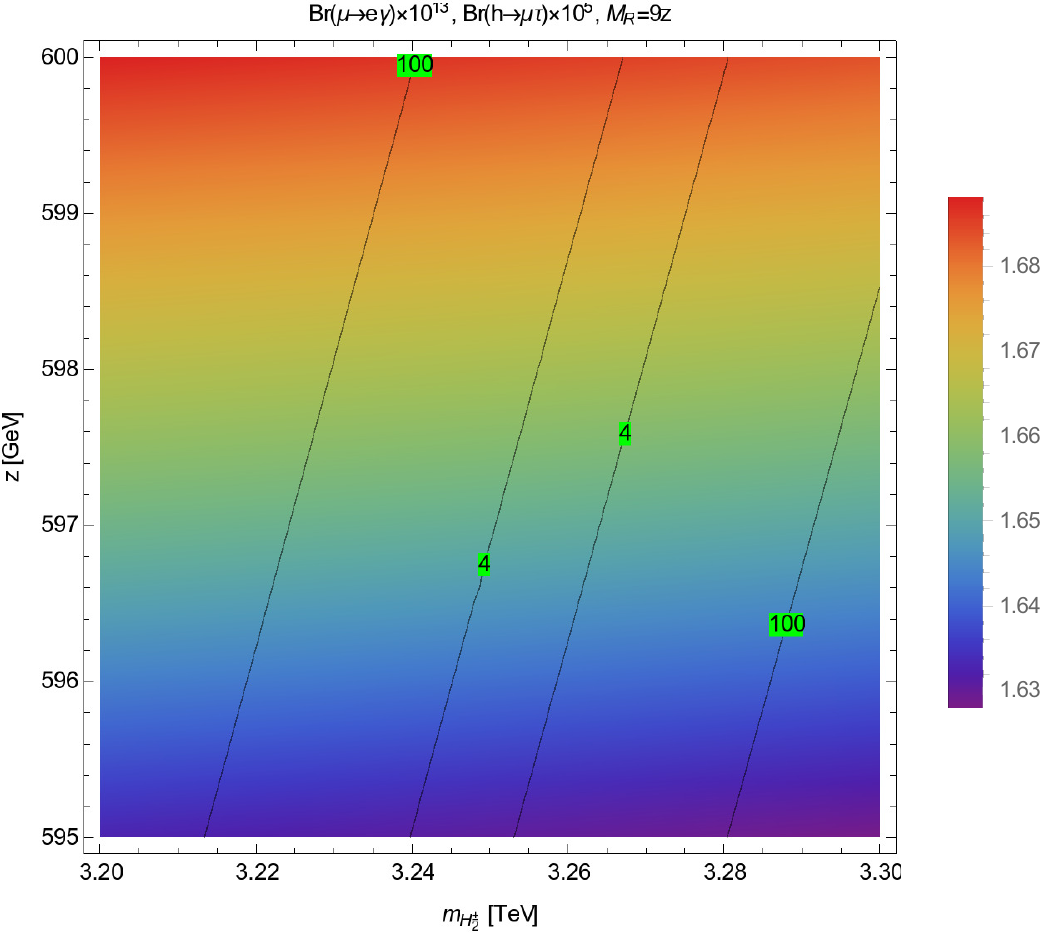}\\
	\end{tabular}
	\caption{Density plots of Br$(h^0_1\rightarrow \mu\tau)$ and contour plots of Br$(\mu\rightarrow e\gamma)$ (black curves) as functions of $m_{H^{\pm}_2}$ and $z$, with $k=5.5$ (upper) and $k=9$ (lower). }\label{DensityPlot}
\end{figure}

The allowed regions in Fig.~\ref{Fdemc2} are shown more explicitly in Fig.~\ref{DensityPlot}, corresponding to $k=5.5$ and $k=9$.
 Only regions that give a large Br$(h^0_1\rightarrow\mu\tau)$ are mentioned.  They are bounded between two black curves representing the constant value of Br$(\mu\rightarrow e\gamma)\times 10^{13}=4$. Clearly, Br$(h^0_1\rightarrow\mu\tau)$ is sensitive to $z$ and $k$, while it changes slowly with changing values of $m_{H^{\pm}_2}$. In contrast, the suppressed Br$(\mu\rightarrow e\gamma)$ allows narrow regions of the parameter space, where some particular relation between $m_{H^{\pm}_2}$ and $k$ and $z$ is realized. Hence, if these two decay channel are discovered by experiments, depending on their specific values, a relation between heavy neutrino and charged Higgs masses can be determined from the 331ISS framework.

\begin{figure}[ht]
	\centering
	\begin{tabular}{cc}
		\includegraphics[width=7cm]{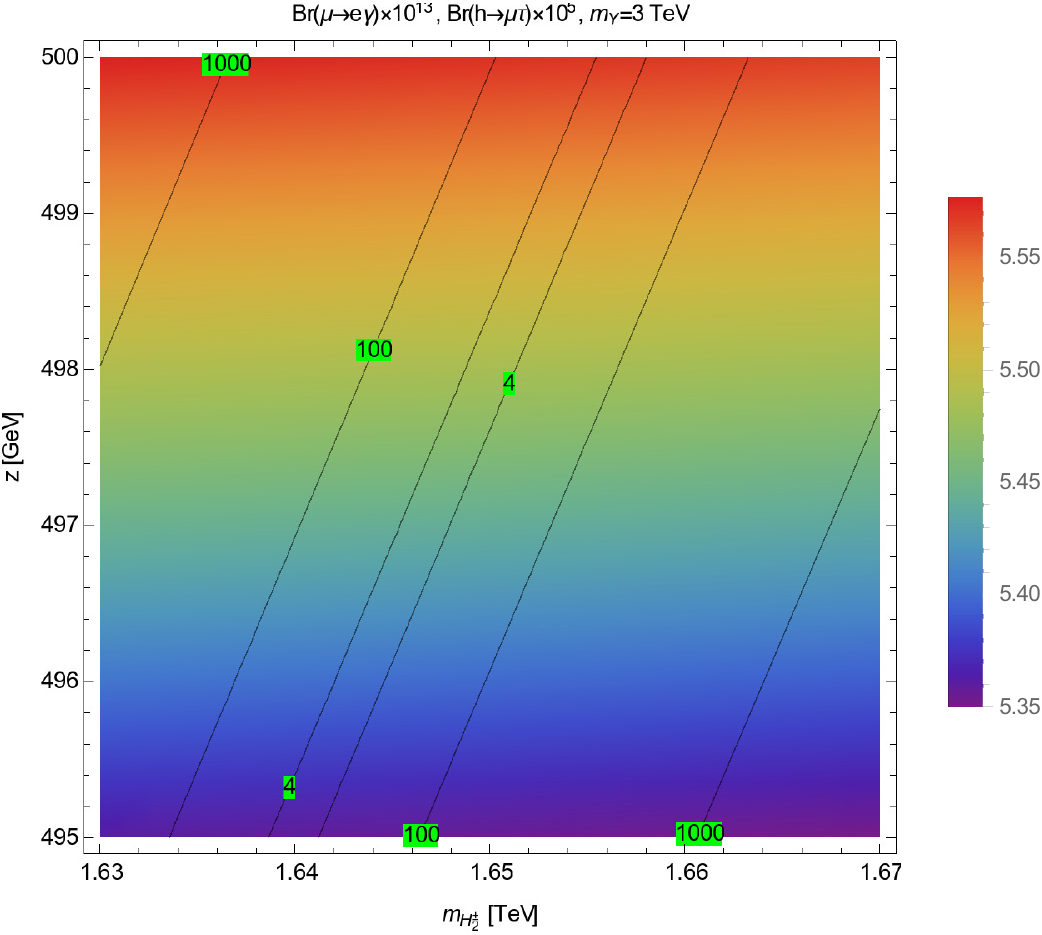}&
		\includegraphics[width=7cm]{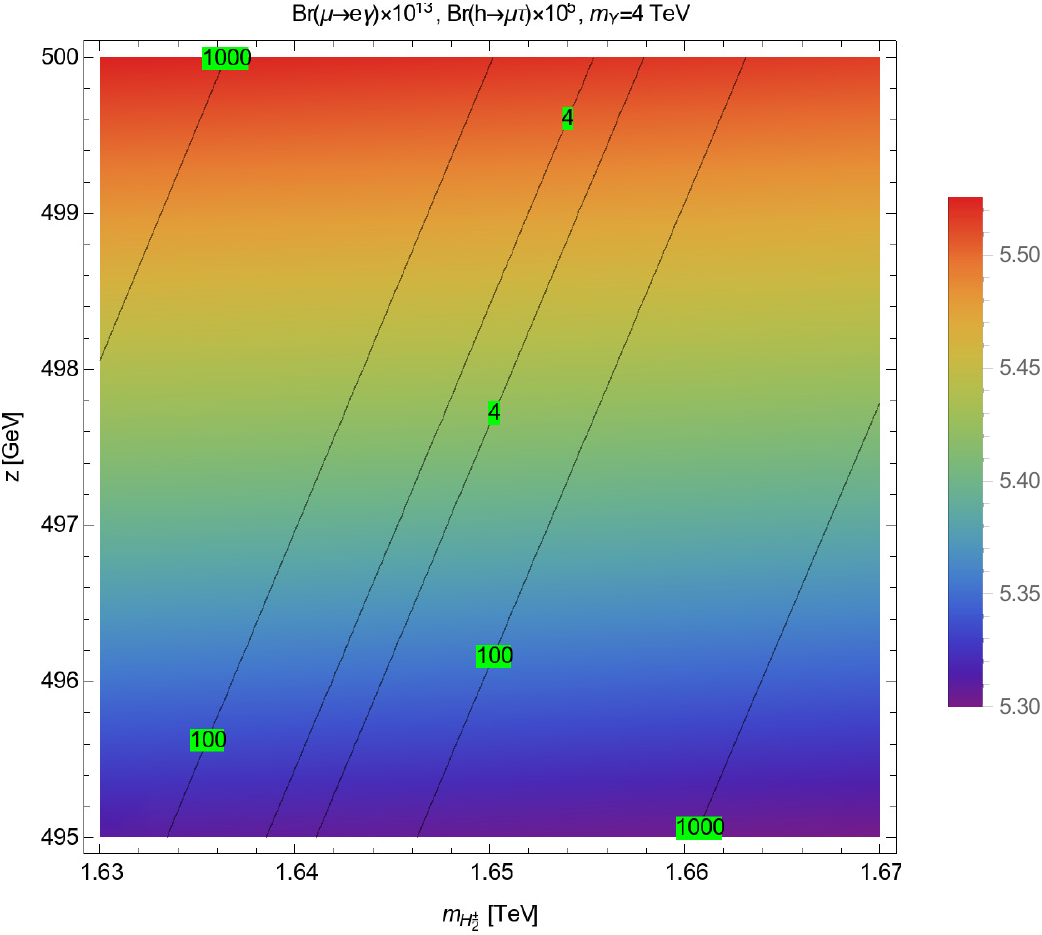}\\
		\includegraphics[width=7cm]{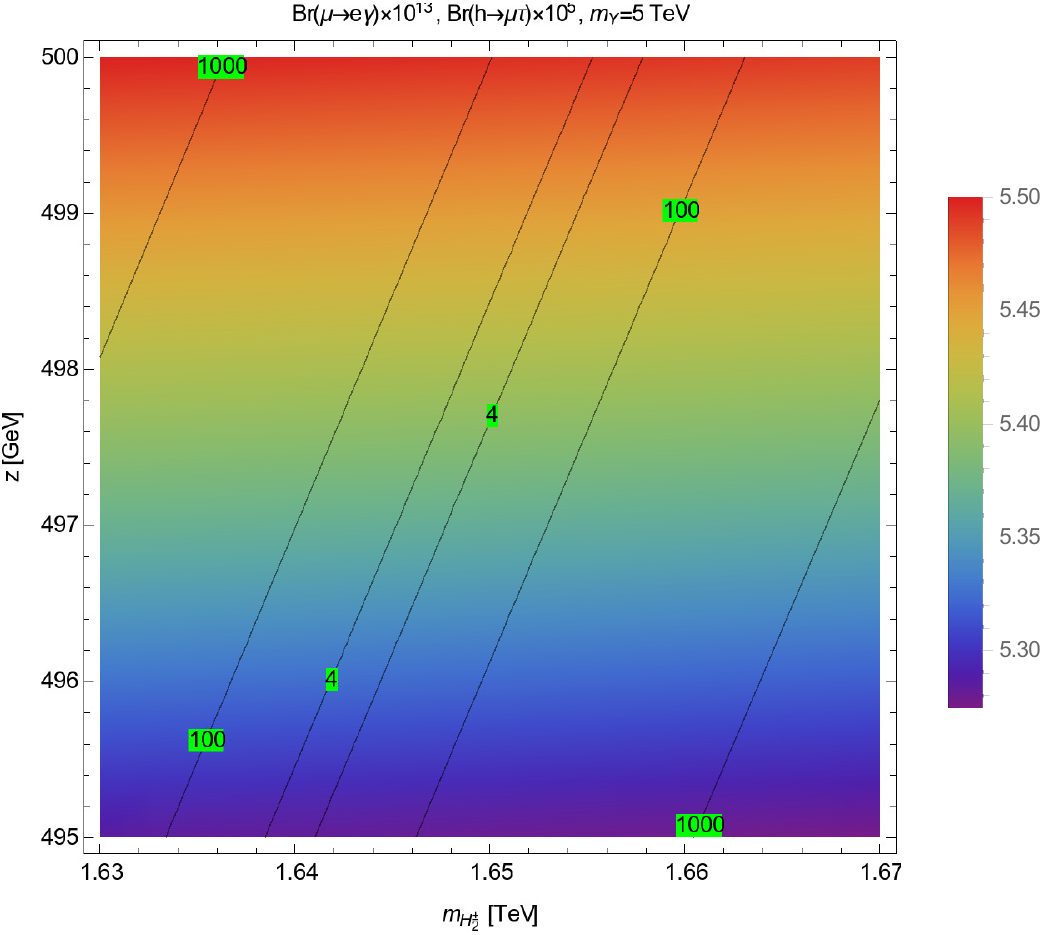}&
		\includegraphics[width=7cm]{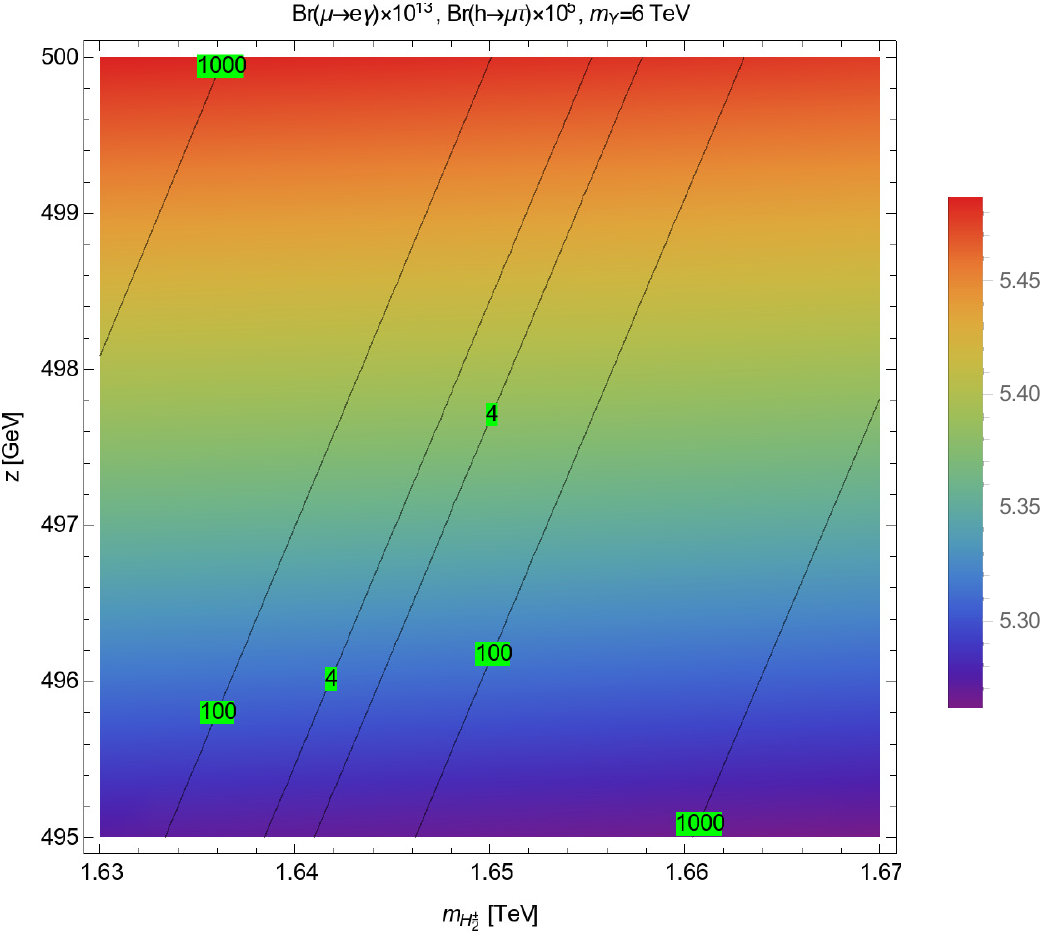}\\
	\end{tabular}
	\caption{Density plots of Br$(h^0_1\rightarrow \mu\tau)$ and contour plots of Br$(\mu\rightarrow e\gamma)$ (black curves) as functions of $m_{H^{\pm}_2}$ and $z$, with $k=5.5$, $z$ around $500$ GeV and different $m_Y$.}\label{DensityPlotk5mY}
\end{figure}

To understand how  Br$(h^0_1\rightarrow\mu\tau)$ depends on the $SU(3)_L$ breaking scale defined by $m_Y$ in this work, four allowed regions corresponding to the four fixed values $m_Y=3,4,5$, and $6$ TeV are illustrated  in  Fig.~\ref{DensityPlotk5mY}.
It can be seen that the branching ratio of LFVHD  depends weakly on  $m_Y$, namely, it decreases slowly with increasing $m_Y$. Hence, studies of LFV decays will give useful information about heavy neutrinos and charged Higgs bosons besides  the phenomenology arising from heavy gauge bosons discussed in many earlier works. More interestingly, this may happen at large $SU(3)_L$ scales which the LHC cannot detect at present.

\section{\label{conclusion} Conclusion}
The 331ISS models seem to be the most interesting among the well-known 331 models because of their rich phenomenology, as indicated in many recent works. This work addressed a more attractive property, namely, the LFVHDs of the SM-like Higgs boson which are being investigated at the LHC. Assuming the absence of the tree-level decays $h^0_1\rightarrow e_ae_b$ and $e_j\rightarrow e_i\gamma$ ($j>i$), the analytical formulas at the one-loop level to calculate these decay rates in the 331ISS model have been introduced.  The divergent cancellation in the total decay amplitudes of $h^0_1\rightarrow e_ae_b$ was shown explicitly. From the numerical investigation, we have indicated that the Br$(h^0_1\rightarrow \mu\tau)$ predicted by the 331ISS model can reach large values of $\mathcal{O}(10^{-5})$. They are even very close to $10^{-4}$, for example, in the special case with $k=5.5$ and  $z\simeq 600$ GeV, which is close to the perturbative limit of the lepton Yukawa couplings. This value is larger than that predicted by the simplest ISS version extended directly from the SM \cite{issE}. New charged Higgs bosons  may give large contributions to both of the decay rates Br$(\mu\rightarrow e\gamma)$ and Br$(h^0_1\rightarrow\mu\tau)$, leading to either constructive or destructive correlations with those from the charged gauge bosons. As a by-product, the recent experimental bound on Br$(\mu\rightarrow e\gamma)$ rules out most of the regions of parameter space with small $k$ and large $z$, except the narrow regions arising from the destructive correlations between contributions of charged Higgs and gauge bosons. We have shown numerically that only these regions give large Br$(h^0_1\rightarrow \mu\tau)>10^{-5}$ when $\mathrm {400}$ GeV $\mathrm{<z<600}$ GeV and $k\le9$ in the case where the Majorana mass matrix $M_R$ is proportional to the identity one. Furthermore,  these large values of Br$(h^0_1\rightarrow\mu\tau)$ depend weakly on the masses of the heavy charged gauge bosons, but they require the heavy neutrino mass scale $M_R$ and $m_{H^{\pm}_2}$ to be a few TeV, which can be detected at current colliders. Besides, Br$(h^0_1\rightarrow \tau e)$ has the same result. In conclusion, large branching ratios of the LFV processes like $h^0_1\rightarrow \mu\tau,e\tau$ will support the 331ISS model and may rule out the original 331RHN model containing only very light exotic neutrinos. Additionally, many properties of heavy neutrinos and charged Higgs bosons in the 331ISS framework may be determined independently at  the $SU(3)_L$ scale.

\section*{Acknowledgments}
This research is funded by the Vietnam National Foundation for Science and Technology Development (NAFOSTED)
under grant number 103.01-2015.33.

\appendix
\section{\label{DeltaLR}Form factors of LFVHDS in the unitary gauge}
In this appendix we list all analytic formulas of one-loop contributions to LFVHDs defined in Eq. (\ref{deLR}). They are written in terms of Passarino-Veltman functions that were defined thoroughly in Refs. \cite{HueNPB16,issthhx}. Using the notations $D_0=k^2-M_0^2+i\delta$, $D_1=(k-p_1)^2-M_{1}^2+i\delta$ and $D_2=(k+p_2)^2-M_2^2+i\delta$, where $\delta$ is an   infinitesimal a positive real quantity, the one-loop integrals and Passarino-Veltman functions needed in this work are
 \bea
B^{(i)}_{0,\mu} &\equiv&\frac{\left(2\pi\mu\right)^{4-D}}{i\pi^2}\int \frac{d^D k\left\{1,k_{\mu}\right\}}{D_0D_i},\quad
B^{(12)}_0 \equiv \frac{\left(2\pi\mu\right)^{4-D}}{i\pi^2}\int \frac{d^D k}{D_1D_2},\crn
C_{0,\mu} &\equiv&   C_{0,\mu}(M_0,M_1,M_2) =\frac{1}{i\pi^2}\int \frac{d^4 k\left\{1,k_{\mu}\right\}}{D_0D_1D_2},\crn
B^{(i)}_{\mu}&=&B^{(i)}_1p_{i\mu}, \hs C_{\mu}=C_1 p_{1\mu} + C_2 p_{2\mu}\nn
\label{scalrInte}\eea
where $i=1,2$.
In addition, $D=4-2\epsilon \leq 4$ is the dimension of the integral,  $M_0,~M_1,~M_2$ are masses of virtual particles in the loop, and $\mu$ is an arbitrary mass parameter introduced via dimensional regularization \cite{DR}. The external momenta of the final leptons shown in Fig.~\ref{hlilj1} satisfy  $p^2_1=m^2_{a},~p^2_2=m^2_{b}$ and $(p_1+p_2)^2=m^2_{h^0_1}$, where $m_{h^0_1}$ is the SM-like Higgs boson mass, and  $m_{a,b}$ are lepton masses. In the limit $m_{a,b}\simeq0$, the analytic formulas for $B^{(i)}_{0,1},\,B^{(12)}_0,\, C_0$, and $C_{1,2}$ were shown in  Refs. \cite{loop,HueNPB16,issthhx}, and hence we will not repeat them here.  These functions are used for our numerical investigation. We stress that they were checked numerically to be well consistent with the exact results computed by LoopTooLS \cite{looptools}, as reported in Ref. \cite{KhiemPTEP16}.

The analytic expressions for $\Delta^{(i)W}_{L,R}\equiv \Delta^{(i)W}_{(ab)L,R}$, where $i$ implies the diagram (i) in Fig. \ref{hlilj1}, are
\bea
\Delta^{(1)W}_{L} &=& \frac{g^3c_{\alpha}  m_{a}}{64\pi^2 m_W^3}\sum_{i=1}^{9}U^{\nu*}_{ai}U^{\nu}_{bi}
\left\{ m_{n_i}^2\left(B^{(1)}_1- B^{(1)}_0- B^{(2)}_0\right) -m_b^2 B^{(2)}_1  +\left(2m_W^2+m^2_{h^0_1}\right)m_{n_i}^2 C_0 \right.\crn &-&\left. \left[2m_W^2\left(2m_W^2+m_{n_i}^2+m_a^2-m_b^2\right) + m_{n_i}^2m_{h^0_1}^2\right] C_1 +
\left[2m_W^2\left(m_a^2-m^2_{h^0_1}\right)+ m_b^2 m^2_{h^0_1}\right]C_2\frac{}{}\right\},\crn
\Delta^{(1)W}_{R}&=& \frac{g^3 c_{\alpha}m_b}{64\pi^2 m_W^3}\sum_{i=1}^{9}U^{\nu*}_{ai}U^{\nu}_{bi}
\left\{ -m_{n_i}^2\left(B^{(2)}_1+B^{(1)}_0+ B^{(2)}_0\right) +m_a^2 B^{(1)}_1  +\left(2m_W^2+m^2_{h^0_1}\right)m_{n_i}^2 C_0 \right.\crn &-&\left.
\left[2m_W^2\left(m_b^2-m^2_{h}\right)+ m_a^2 m^2_{h^0_1}\right]C_1 + \left[2m_W^2\left(2m_W^2+m_{n_i}^2-m_a^2+m_b^2\right) + m_{n_i}^2m_{h^0_1}^2\right] C_2 \frac{}{}\right\},	
\label{de1wL}\crn
\Delta^{(5)W}_{L} &=& \frac{g^3c_{\alpha} m_a}{64\pi^2   m_W^3}\sum_{i,j=1}^{9}U^{\nu*}_{ai}U^{\nu}_{bj} \left\{\lambda^{0*}_{ij}m_{n_j}\left[B^{(12)}_0-m_W^2C_0+\left(2 m_W^2+m_{n_i}^2-m_a^2\right)C_1\right]\right. \crn
&&\hspace{4cm}\left.+\lambda^{0}_{ij}m_{n_i}\left[B^{(1)}_1+\left(2 m_W^2+m_{n_j}^2-m_b^2\right)C_1\right]\right\},\crn
\Delta^{(5)W}_{R} &=&\frac{g^3 c_{\alpha}m_b}{64\pi^2 m_W^3}\sum_{i=1}^{9}U^{\nu*}_{ai}U^{\nu}_{bj} \left\{\lambda^{0}_{ij}m_{n_i}\left[B^{(12)}_0-m_W^2C_0-\left(2 m_W^2+m_{n_j}^2-m_b^2\right)C_2\right]\right. \crn
&&\hspace{3.7cm}-\left.\lambda^{0*}_{ij}m_{n_j}\left[B^{(2)}_1+\left(2 m_W^2+m_{n_i}^2-m_a^2\right)C_2\right]\right\},\crn
\Delta^{(7+8)W}_{L} &=& \frac{g^3m_am_b^2c_{\alpha}}{64\pi^2m^3_W(m_a^2-m_b^2)} \sum_{i=1}^{9}U^{\nu*}_{ai}U^{\nu}_{bi}\left[  2m_{n_i}^2\left(B^{(1)}_0-B^{(2)}_0\right) \right. \crn&-&\left. \left(2 m_W^2 +m_{n_i}^2\right) \left(B^{(1)}_1 +B^{(2)}_1 \right)- m_a^2 B^{(1)}_1 -m_b^2 B^{(1)}_2 \right],  
\crn
\Delta^{(7+8)W}_{R} &=&\frac{m_a}{m_b}\Delta^{(7+8)W}_{L}.
\label{d78LW}\eea
 Defining $\Delta^{(i)Y}_{L,R}= \Delta^{(i)YH^{\pm}_1}_{(ab)L,R}+ \Delta^{(i)YH^{\pm}_2}_{(ab)L,R}$ with $i=4,6,9,10$, the analytic expressions for $\Delta^{(i)Y}_{L,R}\equiv \Delta^{(i)Y}_{(ab)L,R}$ are
\bea \Delta^{(1)Y}_{L} &=&-\frac{g^3  m_{a}\left(\sqrt{2}s_{\alpha}c_{\theta}-c_{\alpha}s_{\theta}\right)}{64 \sqrt 2 \pi^2 m_Y^3}\sum_{i=1}^{9}U^{\nu*}_{(a+3)i}U^{\nu}_{(b+3)i}
\left\{ m_{n_i}^2\left(B^{(1)}_1- B^{(1)}_0-  B^{(2)}_0\right) -m_b^2 B^{(2)}_1\right.\crn
 &+&\left.\left(2m_Y^2+m^2_{h^0_1}\right)m_{n_i}^2 C_0 -  \left[2m_Y^2\left(2m_Y^2+m_{n_i}^2+m_a^2-m_b^2\right)  + m_{n_i}^2m_{h^0_1}^2\right] C_1 \right.\crn
&+&\left.
\left[2m_Y^2\left(m_a^2-m^2_{h^0_1}\right)+ m_b^2 m^2_{h^0_1}\right]C_2\frac{}{}\right\},\crn
\Delta^{(1)Y}_{R} &=& -\frac{g^3 m_{b}\left(\sqrt{2}s_{\alpha}c_{\theta}-c_{\alpha}s_{\theta}\right)}{64 \sqrt 2 \pi^2 m_Y^3}\sum_{i=1}^{9}U^{\nu*}_{(a+3)i}U^{\nu}_{(b+3)i}
\left\{-m_{n_i}^2\left(B^{(2)}_1+B^{(1)}_0+ B^{(2)}_0\right) +m_a^2 B^{(1)}_1 \right.\crn
&+&\left.\left(2m_Y^2+m^2_{h^0_1}\right)m_{n_i}^2 C_0-\left[2m_Y^2\left(m_b^2-m^2_{h^0_1}\right)+ m_a^2 m^2_{h^0_1}\right]C_1  \right.\crn &+&\left.
 \left[2m_Y^2\left(2m_Y^2+m_{n_i}^2-m_a^2+m_b^2\right) + m_{n_i}^2m_{h^0_1}^2\right] C_2 \frac{}{}\right\},\crn
 \Delta^{(2)Y}_{L} &=&\frac{g^3 m_{a} c_{\theta}\left( c_\al c_\theta + \sqrt 2 s_\al s_\theta \right) }{64 \pi^2 m_Wm_Y^2}\sum_{i=1}^{9}U^{\nu*}_{(a+3)i}\crn
 &\times&\left\{\lambda^{L,1}_{bi}m_{n_i}\left[ B^{(1)}_0-B^{(1)}_1+\left(m_Y^2+m_{H^\pm_1}^2-m_{h^0_1}^2\right)C_0+\left(m_Y^2-m_{H^\pm_1}^2+m_{h^0_1}^2\right)C_1\right]\right.\crn
 && +\left. \lambda^{R,1}_{bi}m_{b}\left[ 2m_Y^2C_1-\left(m_Y^2+m_{H^\pm_1}^2-m_{h^0_1}^2\right)C_2\right]\right\},\crn
\Delta^{(2)Y}_{R} &=&\frac{g^3 c_{\theta}\left(c_{\alpha}c_{\theta} +\sqrt{2}s_{\alpha}s_{\theta}\right)}{64 \pi^2 m_Wm_Y^2}\sum_{i=1}^{9}U^{\nu*}_{(a+3)i}\crn
&\times&\left\{\lambda^{L,1}_{bi}m_bm_{n_i}\left[ -2m_Y^2C_0-\left(m_Y^2-m_{H^\pm_1}^2+m_{h^0_1}^2\right)C_2\right]\right.\crn
&&+
\left. \lambda^{R,1}_{bi}\left[-m_{n_i}^2 B^{(1)}_0+m_a^2B^{(1)}_1+m_{n_i}^2\left(m_Y^2-m_{H^\pm_1}^2+m_{h^0_1}^2\right)C_0\right.\right.\crn
&&+\left.\left.\left[ 2m_Y^2\left(m_{h^0_1}^2-m_b^2\right)- m_a^2\left(m_Y^2-m_{H^\pm_1}^2+m_{h^0_1}^2\right)\right]C_1 +2 m_b^2m_Y^2C_2\right]\right\},\crn
 \Delta^{(3)Y}_{L} &=&\frac{g^3 c_{\theta}\left(c_{\alpha}c_{\theta} +\sqrt{2}s_{\alpha}s_{\theta}\right)}{64\pi^2 m_Wm_Y^2}\sum_{i=1}^{9}U^{\nu}_{(b+3)i}\crn
 &\times&\left\{\lambda^{L,1*}_{ai}m_am_{n_i}\left[ -2m_Y^2C_0+\left(m_Y^2-m_{H^\pm_1}^2+m_{h^0_1}^2\right)C_1\right]\right.\crn
 &&+
 \left. \lambda^{R,1*}_{ai}\left[-m_{n_i}^2 B^{(2)}_0-m_b^2B^{(2)}_1+m_{n_i}^2\left(m_Y^2-m_{H^\pm_1}^2+m_{h^0_1}^2\right)C_0\right.\right.\crn
 &&-\left.\left. 2m_a^2m_Y^2C_1-\left[ 2m_Y^2\left(m_{h^0_1}^2-m_a^2\right)- m_b^2\left(m_Y^2-m_{H^\pm_1}^2+m_{h^0_1}^2\right)\right]C_2\right]\right\},\crn
\Delta^{(3)Y}_{R} &=&\frac{g^3 m_b c_{\theta}\left(c_{\alpha}c_{\theta} +\sqrt{2}s_{\alpha}s_{\theta}\right)}{64\pi^2 m_Wm_Y^2}\sum_{i=1}^{9}U^{\nu}_{(b+3)i}\crn
&\times&\left\{\lambda^{L,1*}_{ai}m_{n_i}\left[ B^{(2)}_0+B^{(2)}_1\right.\right.\crn
&&\left.\left.+\left(m_Y^2+m_{H^\pm_1}^2-m_{h^0_1}^2\right)C_0-\left(m_Y^2-m_{H^\pm_1}^2+m_{h^0_1}^2\right)C_2\right]\right.\crn
&& +\left. \lambda^{R,1*}_{ai}m_{a}\left[  \left(m_Y^2+m_{H^\pm_1}^2-m_{h^0_1}^2\right)C_1-2m_Y^2C_2\right]\right\},\crn
\Delta^{(4)YH^{\pm}_{k}}_{L}&=&\frac{g^2\lambda^{\pm}_{H_k}f_k}{16\pi^2  m_W^2}\sum_{i=1}^{9}\left[-\lambda^{R,k*}_{ai}\lambda^{L,k}_{bi}m_{n_i}C_0 -\lambda^{L,k*}_{ai}\lambda^{L,k}_{bi}m_{a}C_1 +\lambda^{R,k*}_{ai}\lambda^{R,k}_{bi}m_{b}C_2 \right],\crn
\Delta^{(4)YH^{\pm}_{k}}_{R} &=&\frac{g^2\lambda^{\pm}_{H_k}f_k}{16\pi^2 m_W^2}\sum_{i=1}^{9}\left[-\lambda^{L,k*}_{ai}\lambda^{R,k}_{bi}m_{n_i}C_0 -\lambda^{R,k*}_{ai}\lambda^{R,k}_{bi}m_{a}C_1 +\lambda^{L,k*}_{ai}\lambda^{L,k}_{bi}m_{b}C_2 \right],\crn
\Delta^{(5)Y}_{L} &=&\frac{g^3c_{\alpha} m_a}{64\pi^2  m_Wm_Y^2}\crn
&\times&\sum_{i,j=1}^{9}U^{\nu*}_{(a+3)i}U^{\nu}_{(b+3)j} \left\{\lambda^{0*}_{ij}m_{n_j}\left[B^{(12)}_0-m_Y^2C_0+\left(2 m_Y^2+m_{n_i}^2-m_a^2\right)C_1\right]\right. \crn
&&\hspace{3.5cm}+\left.\lambda^{0}_{ij}m_{n_i}\left[B^{(1)}_1+\left(2 m_Y^2+m_{n_j}^2-m_b^2\right)C_1\right]\right\},\crn
\Delta^{(5)Y}_{R} &=&\frac{g^3c_{\alpha} m_b}{64\pi^2 m_Wm_Y^2}\crn
&\times&\sum_{i,j=1}^{9}U^{\nu*}_{(a+3)i}U^{\nu}_{(b+3)j} \left\{\lambda^{0}_{ij}m_{n_i}\left[B^{(12)}_0-m_Y^2C_0-\left(2 m_Y^2+m_{n_j}^2-m_b^2\right)C_2\right]\right. \crn
&&\hspace{3.cm}-\left.\lambda^{0*}_{ij}m_{n_j}\left[B^{(2)}_1+\left(2 m_Y^2+m_{n_i}^2-m_a^2\right)C_2\right]\right\},\crn
\Delta^{(6)YH^{\pm}_k}_{L} &=&   -\frac{g^3c_{\alpha}f_k}{32\pi^2  m_W^3}\sum_{i,j=1}^{9}\left\{\lambda^{0*}_{ij}\left[\lambda^{R,k*}_{ai}\lambda^{L,k}_{bj}\left(B^{(12)}_0+m_{H^\pm_k}^2C_0 -m_a^2 C_1+m_b^2C_2\right)\right.\right.\crn
&+&\left.\left. \lambda^{R,k*}_{ai}\lambda^{R,k}_{bj}m_bm_{n_j}C_2 -\lambda^{L,k*}_{ai}\lambda^{L,k}_{bj}m_am_{n_i}C_1 \right]\right. \crn
&+&\left.  \lambda^{0}_{ij}\left[\lambda^{R,k*}_{ai}\lambda^{L,k}_{bj}m_{n_i}m_{n_j}C_0 +\lambda^{R,k*}_{ai}\lambda^{R,k}_{bj}m_{n_i}m_{b}(C_0+C_2)\right.\right.\crn
&+&\left.\left.\lambda^{L,k*}_{ai}\lambda^{L,k}_{bj}m_{a}m_{n_j}(C_0-C_1)+ \lambda^{L,k*}_{ai}\lambda^{R,k}_{bj}m_{a}m_{b}(C_0-C_1+C_2) \right]\frac{}{}\right\},\crn
\Delta^{(6)YH^{\pm}_k}_{R} &=&-\frac{g^3c_{\alpha}f_k}{32\pi^2  m_W^3} \sum_{i,j=1}^{9}\left\{\lambda^{0}_{ij}\left[\lambda^{L,k*}_{ai}\lambda^{R,k}_{bj}\left(B^{(12)}_0+m_{H^\pm_k}^2C_0 -m_a^2 C_1+m_b^2C_2\right)\right.\right.\crn
&+&\left.\left.\lambda^{L,k*}_{ai}\lambda^{L,k}_{bj}m_bm_{n_j}C_2-  \lambda^{R,k*}_{ai}\lambda^{R,k}_{bj}m_am_{n_i}C_1 \right]\right.\crn
&+&\left.
 \lambda^{0*}_{ij}\left[\lambda^{L,k*}_{ai}\lambda^{R,k}_{bj}m_{n_i}m_{n_j}C_0 +\lambda^{L,k*}_{ai}\lambda^{L,k}_{bj}m_{n_i}m_{b}(C_0+C_2)\right.\right.\crn
&+&\left.\left.\lambda^{R,k*}_{ai}\lambda^{R,k}_{bj}m_{a}m_{n_j}(C_0-C_1)+ \lambda^{R,k*}_{ai}\lambda^{L,k}_{bj}m_{a}m_{b}(C_0-C_1+C_2) \right]\frac{}{}\right\},\crn
\Delta^{(7+8)Y}_{L} &=&\frac{g^3m_am_b^2c_{\alpha}}{64\pi^2m_Wm_Y^2(m_a^2-m_b^2)} \sum_{i=1}^{9}U^{\nu*}_{(a+3)i}U^{\nu}_{(b+3)i} \crn &\times&\left[  2m_{n_i}^2\left(B^{(1)}_0-B^{(2)}_0\right)-\left(2 m_Y^2 +m_{n_i}^2\right) \left(B^{(1)}_1 +B^{(2)}_1 \right) - m_a^2 B^{(1)}_1 -m_b^2 B^{(2)}_1 \right],  \label{d78YL}\crn
\Delta^{(7+8)Y}_{R} &=&\frac{m_a}{m_b}\Delta^{(7+8)Y}_{L},\crn
\Delta^{(9+10)YH^{\pm}_k}_{L}  &=&-\frac{g^3c_{\alpha}f_k}{32\pi^2m_W^3\left(m_a^2-m_b^2\right)} \crn
&\times&\sum_{i=1}^{9}\left[ m_am_bm_{n_i}  \lambda^{L,k*}_{ai}\lambda^{R,k}_{bi}\left(B^{(1)}_0-B^{(2)}_0\right)+ m_{n_i}
\lambda^{R,k*}_{ai}\lambda^{L,k}_{bi}\left(m^2_bB^{(1)}_0-m^2_aB^{(2)}_0\right)\right.\crn
&&\left.+ m_{a}m_b \left(\lambda^{L,k*}_{ai}\lambda^{L,k}_{bi}m_b + \lambda^{R,k*}_{ai}\lambda^{R,k}_{bi}m_a\right)\left(B^{(1)}_1+ B^{(2)}_1\right)\right],\crn
\Delta^{(9+10)YH^{\pm}_k}_{R}  &=&-\frac{g^3c_{\alpha}f_k}{32\pi^2m_W^3\left(m_a^2-m_b^2\right)} \crn
&\times&\sum_{i=1}^{9}\left[ m_am_bm_{n_i}  \lambda^{R,k*}_{ai}\lambda^{L,k}_{bi}\left(B^{(1)}_0-B^{(2)}_0\right)+ m_{n_i}
\lambda^{L,k*}_{ai}\lambda^{R,k}_{bi}\left(m^2_bB^{(1)}_0-m^2_aB^{(2)}_0\right)\right.\crn
&&\left.+ m_{a}m_b \left(\lambda^{R,k*}_{ai}\lambda^{R,k}_{bi}m_b + \lambda^{L,k*}_{ai}\lambda^{L,k}_{bi}m_a\right)\left(B^{(1)}_1+ B^{(2)}_1\right)\right],
\label{dltaLRY}\eea
where $f_{1}=c^2_{\theta}$ and $f_{2}=1/2$.
The details to  derive the  expressions in Eq. (\ref{dltaLRY})  are  the same as those shown in Refs. \cite{issthhx,HueNPB16}, and hence we do not present them in this work. We note that  the scalar functions $\Delta^{(1)W}_{L,R}$  and $\Delta^{(1,2,3)Y}_{L,R}$ include  parts that do not depend on $m_{n_i}$, and therefore they vanish because of the Glashow-Iliopoulos-Maiani mechanism.  They are ignored in Eqs. (\ref{d78LW}) and  (\ref{dltaLRY}).

The divergent cancellation in the total $\Delta_{L,R}$ is shown as follows.  The divergent parts only contain  $B$ functions: div$B^{(1)}_0=$div$B^{(2)}_0=$div$B^{(12)}_0=2$div$B^{(1)}_1=-2$ div$B^{(2)}_1=\Delta_{\epsilon}$. Ignoring the common factor of $g^3/(64\pi^2m_W^3)$ and using $1/m_Y=\sqrt{2}s_{\theta}/m_W$, the divergent parts of $\Delta_L$ derived from Eq. (\ref{dltaLRY}) are
\bea \mathrm{div}\left[\Delta^{(1)W}_L\right]&=&m_a\Delta_{\epsilon}\times \left( -\frac{3c_{\al}}{2}\right) \sum_{i=1}^9 U^{\nu*}_{ai}U^{\nu}_{bi} m^2_{n_i},\crn
\mathrm{div}\left[\Delta^{(5)W}_L\right]&=&m_a \Delta_{\epsilon}\times c_{\al}\sum_{i,j=1}^9 U^{\nu*}_{ai}U^{\nu}_{bj}\left(\lambda^{0*}_{ij} m_{n_j}+\frac{1}{2}\lambda^{0}_{ij} m_{n_i}\right),\crn
\mathrm{div}\left[\Delta^{(7+8)W}_L\right]&=&\mathrm{div}\left[\Delta^{(4)Y}_L\right]= \mathrm{div}\left[\Delta^{(7+8)Y}_L\right]=0,\crn
\mathrm{div}\left[\Delta^{(1)Y}_L\right]&=&m_a \Delta_{\epsilon}\times 3s_{\theta}^3 \left(\sqrt 2 s_\al c_\theta - c_\al s_\theta \right) \sum_{i=1}^9 U^{\nu*}_{(a+3)i}U^{\nu}_{(b+3)i} m^2_{n_i},\crn
\mathrm{div}\left[\Delta^{(2)Y}_L\right]&=&m_a\Delta_{\epsilon}\times s_{\theta}^2 c_\theta \left( c_\al c_\theta + \sqrt 2 s_\al s_\theta \right)\sum_{i=1}^9 U^{\nu*}_{(a+3)i}\lambda^{L,1}_{bi} m_{n_i},\crn
\mathrm{div}\left[\Delta^{(3)Y}_L\right]&=&m_a\Delta_{\epsilon}\times  \left[-2s_{\theta}^2 c_\theta  \left( c_\al c_\theta + \sqrt 2 s_\al s_\theta \right)\right] \sum_{i=1}^9 U^{\nu*}_{(a+3)i}U^{\nu}_{(b+3)i} m^2_{n_i},\crn
\mathrm{div}\left[\Delta^{(5)Y}_L\right]&=&m_a \Delta_{\epsilon}\times  2s_{\theta}^2c_{\al} \sum_{i,j=1}^9 U^{\nu*}_{(a+3)i}U^{\nu}_{(b+3)j} \left(\lambda^{0*}_{ij} m_{n_j}+\frac{1}{2}\lambda^{0}_{ij} m_{n_i}\right),\crn
\mathrm{div}\left[\Delta^{(6)YH_1^\pm}_L\right]&=&m_a\Delta_{\epsilon}\times \left(-2c_{\alpha}c_{\theta}^2 \right) \sum_{i,j=1}^9 U^{\nu*}_{(a+3)i}\lambda^{0*}_{ij}\lambda^{L,1}_{bj},\crn
\mathrm{div}\left[\Delta^{(6)YH_2^\pm}_L\right]&=&m_a\Delta_{\epsilon}\times \left(-c_{\alpha}\right) \sum_{i,j=1}^9 U^{\nu*}_{ai}\lambda^{0*}_{ij}\lambda^{L,2}_{bj},\crn
\mathrm{div}\left[\Delta^{(9+10)YH_1^\pm}_L\right]&=&m_a\Delta_{\epsilon}\times \left( 2c_{\alpha}c_{\theta}^2\right) \sum_{i=1}^9 U^{\nu*}_{(a+3)i}\lambda^{L,1}_{bi} m_{n_i}, \crn
\mathrm{div}\left[\Delta^{(9+10)YH_2^\pm}_L\right]&=&m_a\Delta_{\epsilon}\times  c_{\alpha} \sum_{i=1}^9 U^{\nu*}_{ai}\lambda^{L,2}_{bi} m_{n_i},
\label{divdeli}\eea
Using the equalities $M^{\nu}=U^{\nu*}\hat{M^{\nu}}U^{\nu\dagger}$ and   Eq. (\ref{fmnu2}), we can prove that
\bea \mathrm{div}\left[\Delta^{(1)W}_{L,R}\right] &\sim &\sum_{i=1}^9U^{\nu*}_{ai}U^{\nu}_{bi}m^2_{n_i}=\left[U^{\nu}(\hat{M}^{\nu})^2U^{\nu\dagger}\right]_{ba}=(m_D^*m_D^T)_{ba}\crn
&=&(m_D^\dagger m_D)_{ba},
\crn
\mathrm{div}\left[\Delta^{(5)W}_{L,R}\right] &\sim & \sum_{i,j=1}^9U^{\nu*}_{ai}U^{\nu}_{bj}\lambda^{0*}_{ij}m_{n_j},\, \sum_{i,j=1}^9U^{\nu*}_{ai}U^{\nu}_{bj}\lambda^{0}_{ij}m_{n_i}=(m_D^*m_D^T)_{ba}\crn
&=&(m_D^\dagger m_D)_{ba},
\crn
\mathrm{div}\left[\Delta^{(1,3)Y}_{L,R}\right] &\sim& \sum_{i=1}^9U^{\nu*}_{(a+3)i}U^{\nu}_{(b+3)i}m^2_{n_i}=\left[U^{\nu}(\hat{M}^{\nu})^2U^{\nu\dagger}\right]_{(b+3)(a+3)}\crn &=&(m_D^\dagger m_D+ M_R^*M_R^T)_{ba},
\crn
\mathrm{div}\left[\Delta^{(2)Y,\, (9+10)YH^\pm_1}_{L,R}\right] &\sim &\sum_{i=1}^9U^{\nu*}_{(a+3)i}\lambda^{L,1}_{bi}m_{n_i}=(m_D^*m_D)_{ba}+  t^2_{\theta}(M^*_RM^T_R)_{ba}\crn
&=&-(m_D^{\dagger}m_D)_{ba}+ t^2_{\theta}(M^*_RM^T_R)_{ba},
\crn
\mathrm{div}\left[\Delta^{(5)Y}_{L,R}\right] &\sim & \sum_{i,j=1}^9U^{\nu*}_{(a+3)i}U^{\nu}_{(b+3)j}\lambda^{0*}_{ij}m_{n_j},\, \sum_{i,j=1}^9U^{\nu*}_{(a+3)i}U^{\nu}_{(b+3)j}\lambda^{0}_{ij}m_{n_i}\crn
&\sim&(m_D^\dagger m_D)_{ba}-\sqrt2 t_{\alpha}t_{\theta}(M^*_RM^T_R)_{ba},\crn
\mathrm{div}\left[\Delta^{(6)YH^\pm_1}_{L,R}\right] &\sim &  \sum_{i,j=1}^9U^{\nu*}_{(a+3)i}\lambda^{0*}_{ij}\lambda^{L,1}_{bj}=(m_D^* m_D)_{ba}- \sqrt 2 t_{\alpha}t_{\theta}^3 (M^*_RM^T_R)_{ba}\crn
 &=&-(m_D^\dagger m_D)_{ba}-\sqrt2 t_{\alpha}t_{\theta}^3(M^*_RM^T_R)_{ba},\crn
 \mathrm{div}\left[\Delta^{(6)YH^\pm_2}_{L,R}\right] &\sim &  \sum_{i,j=1}^9U^{\nu*}_{ai}\lambda^{0*}_{ij}\lambda^{L,2}_{bj}=-(m_D^\dagger m_D)_{ba},\crn
 \mathrm{div}\left[\Delta^{(9+10)YH^\pm_2}_{L,R}\right] &\sim &  \sum_{i=1}^9U^{\nu*}_{ai}\lambda^{L,2}_{bi}m_{n_i}=-(m_D^\dagger m_D)_{ba},
\label{usex}\eea
where we have used the antisymmetric property of $m_D$: $m_D^T=-m_D$.   From this, it can be seen that  $\mathrm{div}\left[\Delta^{(1)W}_L\right]+\mathrm{div}\left[\Delta^{(5)W}_L\right]=\mathrm{div}\left[\Delta^{(6)YH^\pm_2}_L\right]+\mathrm{div}\left[\Delta^{(9+10)YH^\pm_2}_L\right]=0$. The sum of  the  remaining divergent parts  is
\begin{align}
& \mathrm{div}\left[\Delta^{(1+2+3+5)Y}_L+\Delta^{(6+9+10)YH^\pm_1}_L \right]\crn
&\sim (m_D^\dagger m_D)_{ba} \left\{\sqrt{2}s_{\alpha}s^2_{\theta}c_{\theta}(3-1-2) + c_{\alpha}\left[s^2_{\theta}(-3s^2_{\theta}-c^2_{\theta}-2c^2_{\theta}+3) +2 s^2_{\theta}-2s^2_{\theta} \right]\right\}\crn
&+ (M^*_RM^T_R)_{ba}\left[ \sqrt{2}s_{\alpha}\frac{s^2_{\theta}}{c_{\theta}} \left(3c^2_{\theta}+s^2_{\theta}-2c^2_{\theta}-3+2\right)+ c_{\alpha}s^2_{\theta}\left(-3s^2_{\theta}+s^2_{\theta}-2c^2_{\theta}+2\right)\right]\crn
&=0. \label{sdiv}
\end{align}
Finally, the proof of the divergent cancellation in $\Delta_R$ is exactly the same as that in $\Delta_L$.

\end{document}